\documentclass{aa}
\usepackage{graphicx}
\usepackage{txfonts}
\newcommand{\gro}  {GRO\,J2058+42}
\newcommand{\ha}  {H$\alpha$}
\newcommand{\ew}  {EW(H$\alpha$)}
\newcommand{\ergs}  {erg s$^{-1}$}

\def\simless{\mathbin{\lower 3pt\hbox
     {$\rlap{\raise 5pt\hbox{$\char'074$}}\mathchar"7218$}}}   %< or of order
\def\simmore{\mathbin{\lower 3pt\hbox
     {$\rlap{\raise 5pt\hbox{$\char'076$}}\mathchar"7218$}}}   %> or of order

\def\msun{~{\rm M}_\odot}
\def\rsun{~{\rm R}_\odot}

\begin{document}

   \title{Long-term optical variability of the Be/X-ray binary GRO J2058+42}

   \subtitle{}
  \author{P. Reig \inst{1,2}
        \and
        A. Tzouvanou \inst{2}
        \and
        D. Blinov \inst{1,2}
        \and
        V. Pantoulas\inst{1}
           }

\authorrunning{}
\titlerunning{\gro}

   \offprints{pau@physics.uoc.gr}

   \institute{Institute of Astrophysics, Foundation for Research and
   Technology-Hellas, 71110 Heraklion, Greece 
         \and Physics Department, University of Crete, 71003 Heraklion, Greece
                \email{pau@physics.uoc.gr}
        }

   \date{Received ; accepted}

\abstract
%context
{GRO J2058+42 is a transient hard  X-ray pulsar that occasionally goes into
outburst. The optical counterpart is a poorly studied OB-type companion.} 
%aims
{We investigate the long-term optical variability of the Be/X-ray
binary GRO J2058+42 and the possible connection with periods of enhanced X-ray activity. } 
%methods
{We performed an optical spectroscopic and photometric analysis on data collected during about 18
years. We also present the first optical polarimetric observations of this
source. 
%The optical observables allow us to monitor the state of the Be star's circumstellar disk. 
 }
%results
{
%The optical counterpart to GRO J2058+42 displays quasiperiodic variability in the strength of the
%H$\alpha$ line and  continuum optical emission on timescales of about 9.5 years.
The long-term optical light curves in the $BVRI$ bands and the evolution of the
H$\alpha$ equivalent width display a sinusoidal  pattern with maxima and
minima that repeat every  $\sim$9.5 years. The amplitude of this variability
increases as the wavelength increases from 0.3 mag in the $B$ band to 0.8 in the
$I$ band. The H$\alpha$ equivalent width varied from about $-0.3$ to $-15$ \AA. 
We found a significant decrease in the polarization degree during the
low optical state. The intrinsic polarization
degree changed by $\sim 1$\% from maximum to minimum.
The optical maxima occur near periods of enhanced X-ray activity and
are followed by a drop in the optical emission. Unlike many other Be/X-ray
binaries, \gro\ does not display $V/R$ variability.
 }
%conclusions
{The long-term optical variability agrees with the standard model of a Be/X-ray
binary, where the  circumstellar disk of the Be star grows and dissipates on timescales of 9--10 years. We find that the dissipation of the disk started after a
major X-ray outburst. However, the stability of the H$\alpha$ line shape as a
double-peak profile and the lack of asymmetries suggest the absence of a
warped disk and argue against the presence of a highly distorted disk
during major X-ray outbursts.
}


   \maketitle

\section{Introduction}

GRO\,J2058+40 was discovered as a 198 s X-ray pulsar by BATSE on board the
Compton Gamma-Ray Observatory ({\it CGRO}) during a giant outburst in 1995
September-October \citep{wilson98}. This giant (type II) outburst was followed
by a series of weaker (type I) outbursts, whose intensity peaked at intervals of
about 55 days \citep{wilson98,wilson05}. Because BATSE detected the odd
outbursts, counting from the giant outburst, to be brighter than even outbursts,
it was proposed that \gro\ was undergoing periastron and apastron outburst in a
110 day orbit \citep{wilson98}. However, the all-sky monitor (ASM)
on board the {\it Rossi X-Ray Timing Explorer (RXTE)} did not see such a pattern.
Consequently, the orbital period was established to be 55 days \citep{wilson05}.

\gro\ was X-ray active from 1996 to 2002, but only detected at periastron
passages. Pulsations were not detected during the epoch 2003--2004 
\citep{wilson05}.
\gro\ entered a new period of enhanced X-ray activity in May 2008
\citep{krimm08}. Although the X-ray flux doubled in less than a week, the
luminosity was relatively low, $L_x=1.5\times 10^{36}$ \ergs\ assuming a
distance of 9 kpc \citep{wilson05,reig15}. This is more consistent with  the lower-luminosity type I outbursts. However, because the 2008 X-ray data were
limited to just one {\it Swift}/XRT pointed observation, it is likely that 
the peak of the outburst was missed.

Another giant X-ray outburst occurred in March 2019 with a peak luminosity in
the energy range 3--78 keV of $L_x=5.6 \times 10^{37}$ \ergs\ \citep{kabiraj20},
similar to the discovery luminosity back in 1995.  {\it NuSTAR} and {\it
AstroSat} observations performed during this outburst revealed the possible
presence of a cyclotron line at 10 keV together with some harmonics
\citep{molkov19,mukerjee20}. During {\it AstroSat} observations, a
quasi-periodic oscillation at 0.090 Hz was detected during the decay of the
outburst \citep{mukerjee20}. The {\it NuSTAR} observations only detected the
cyclotron line in a very narrow range of the spin phases of \gro\ and covering
only $\sim10$\% of the entire spin period \citep{molkov19}, but it did not seem
to be present in the pulse-average spectrum \citep{kabiraj20}.

The optical counterpart to GRO\,J2058+40 is an O9.5-B0e IV-V star
\citep{reig05a,wilson05}. The massive companion is a long-term variable system
with photometric and spectroscopic changes on timescales of years
\citep{kiziloglu07b,reig15,reig16}. The underlying B star also displays fast
optical multiperiodic variability that is  attributed to nonradial pulsations
\citep{kiziloglu07b,reig22}. We present the most complete and
detailed study of the optical variability of the BeXB \gro\ performed so far.

\section{Observations and data analysis}

We obtained optical photometry, spectroscopy, and polarimetry of the optical
counterpart to \gro. The observations were made from the 1.3 m telescope at the
Skinakas observatory\footnote{https://skinakas.physics.uoc.gr/en/} (SKO) in
Crete (Greece). Additional spectra were retrieved from the data archive of the 
2.0 m Liverpool Telescope\footnote{https://telescope.livjm.ac.uk/}(LT).

\subsection{Photometry}

Photometric observations with the Johnson-Cousins $B$,
$V$, $R,$ and $I$ filters were made with the 1.3 m telescope of the
Skinakas Observatory. The telescope
was equipped with a  2048$\times$2048 ANDOR CCD with a 13.5 $\mu$m pixel
size. In this configuration, the  plate scale is 0.28$\arcsec$/pixel,
hence providing a field of view of $9.5 \times 9.5$ arcmin$^2$. Standard
stars from the Landolt list \citep{landolt09} were used for the
transformation equations.  The data were reduced in the
standard way using the IRAF tools for aperture photometry\footnote{A User's Guide to CCD
Reductions with IRAF, Philip Massey, February 1997. https://iraf.net/irafdocs} 
\citep{tody86}. After the
standardization process, we finally assigned an error to the calibrated
magnitudes of the target given by the rms of the residuals between the
cataloged and calculated magnitudes of the standard stars. The photometric
magnitudes are given in Table~\ref{photres} (see Appendix~\ref{app}).

\subsection{Polarimetry}

Polarimetric observations were made with the RoboPol photopolarimeter attached
to the focus of the 1.3 m telescope of the Skinakas Observatory.  In the
polarimetry configuration a plate scale of  0.43 $\arcsec$/pixel is achieved
with a 2048$\times$2048 ANDOR CCD with a 13.5 $\mu$m pixel size.  RoboPol is an
imaging photopolarimeter that simultaneously measures the Stokes parameters of
linear polarization  of all sources in the field of view
\citep{king14,ramaprakash19}. RoboPol splits the incident light in two beams,
each half incident on a half-wave retarder followed by a Wollaston prism.  The
fast axis of the half-wave retarder in front of the first prism is rotated by
$67.5^{\circ}$ with respect to the other retarder. Every point in the sky is
thereby projected to four points on the CCD with different polarization state. 
The photon counts in each spot, measured using aperture photometry, were used to
calculate the $U$ and $Q$ parameters of linear polarization.  To optimize the
instrument sensitivity, a mask was placed in the telescope focal plane. The
absence of moving parts allows RoboPol to compute all the Stokes parameters of
linear polarization in one shot. The intrumental polarization and polarization
angle zeropoint were controlled with regular measurements of polarimetric
standards as described in \citet{blinov21}. All observations were corrected for the
instrumental polarization, and uncertainties were propagated accordingly. The
results of the polarimetric observations are given in Table~\ref{polres}
(Appendix~\ref{app}).

\subsection{Spectroscopy}

The 1.3\,m telescope at SKO  was equipped with a
2048$\times$2048 (13.5 micron) pixel ANDOR IKON CCD and a 1302 l~mm$^{-1}$
grating, giving a nominal dispersion of $\sim$0.9 \AA/pixel.  Spectra of
comparison lamps were taken before and after each exposure in order to perform
the wavelength calibration and account for small variations produced by the
tension and flexture of the telescope at different zenith angles during the
night that could affect the wavelength calibration The 2.0 m LT is a fully
robotic telescope at the Roque de Los Muchachos on the Canary Island of La Palma
(Spain). We downloaded and analyzed all the spectra available at the LT data
archive for this source.  These spectra were obtained  with the Fibre-fed
RObotic Dual-beam Optical Spectrograph (FRODOSpec), which is  a dual-beam design
multipurpose integral-field input spectrograph that splits the beam before the
entrance to the individually optimized collimators. We used the
spectra taken with the red arm, which cover the wavelength range 5900--8000 \AA\
with a dispersion of 0.6 \AA/pixel. A Xenon arc exposure was obtained after the
science exposure. The log of the spectroscopic observations is given in  
Table~\ref{specres} (see Appendix~\ref{app}).

The \ha\ line is the prime indicator of the state of the disk. The structure and
size of the disk affect not only the strength of the \ha\ line, but also its
shape. We investigated the evolution of the strength and line
profile over about an 18 year period. We took the equivalent width of the \ha\
line (\ew) as a proxy of the strength of the line. To study the changes
experienced by the shape of the line, we fit the line profile with one or two
Lorentzian functions, depending on whether the line displayed a single-peak or a
double-peak profile. In the case of a split profile, the peak at shorter
wavelength is called the blue or violet peak, and it is denoted by $V$, and the
peak at the longer wavelength is called the red peak and is represented by $R$.

The three fitting parameters are the central wavelength ($\lambda_i$), the
full width at half maximum ($FWHM_i$), and the relative intensity of the peak
($I_i$). Here the subindex $i$ refers to either the blue ($V$) or red peak
($R$). When the line showed a double-peak profile, we obtained the
following quantities: $\Delta V=\lambda_{\rm R}-\lambda_{\rm V}$
and $V/R=\log(I_{\rm V}/I_{\rm R})$, where $I_{\rm V}$ and $I_{\rm R}$ are the
intensity of the blue and red peaks relative to the continuum. For single-peak
profiles, $\Delta V=0$ and $\log(V/R)$ was not defined.

The \ew\ was computed by means of a numerical integration method, the
trapezoidal rule, as implemented in the scipy python package. To ensure an
homogeneous processing, all the spectra were normalized with respect to the
local continuum, which  was rectified to unity by employing a polynomial fit.
Because the definition of the continuum is the main source of uncertainty in the
computation of the equivalent width, we averaged over different selections of the
continuum and different polynomial fits (different grade). A total of 24
measurements were obtained for each spectrum. The error in \ew\ is the standard
deviation of these measurements.

For the line fits, we used the {\tt deblend} task of the SPLAT package
\citep{skoda14} of the Starklink project \citep{currie14}. The fits require the
continuum to be at zero level. Therefore, the best polynomial fit of the
continuum was subtracted from the spectra prior to the Lorentzian fits.

%------------------------------Fig. 1------------------------------------------------
\begin{figure} 
\resizebox{\hsize}{!}{\includegraphics{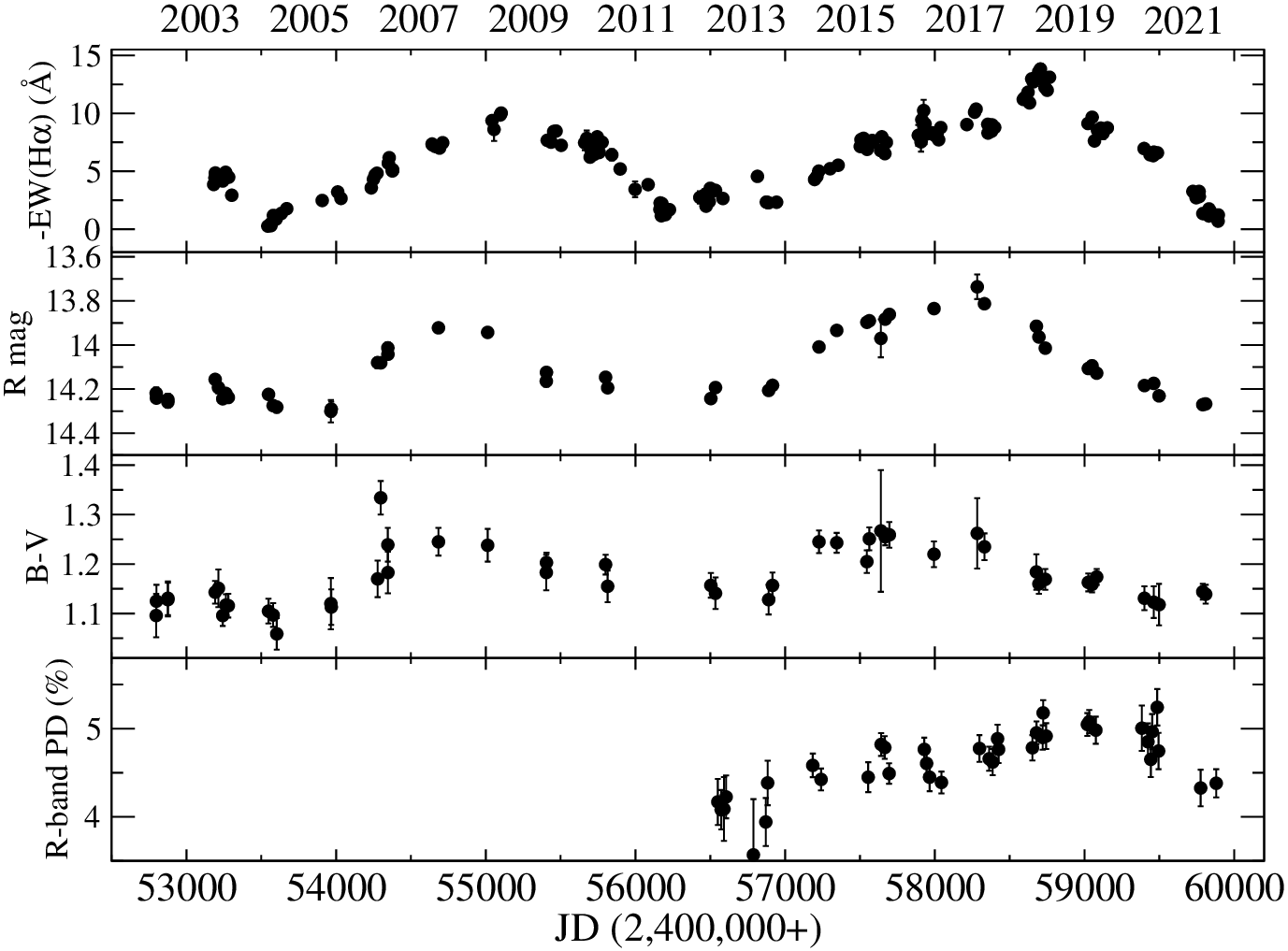}} 
\caption[]{Evolution of the optical observables. From top to bottom, we show the \ha\
equivalent width, continuum emission in the $R$ band, $B-V$ color index, and
polarization degree in the $R$ band (not corrected for ISM polarization).}
 \label{opt}
\end{figure}
%------------------------------------------------------------------------------

\section{Results}

%In this section we present the results of our photometric, spectroscopic, and
%polarimetric analysis.

\subsection{Brightness variability}
\label{res}

Figure~\ref{opt} shows the long-term optical variability of \gro. In general, 
the correlation among all optical observables is good. As the strength of
the line emission increases, the continuum emission and the polarization 
increase.  
\gro\ went through two high and three low optical states during the course of
our observations. 

A closer look at Fig.~\ref{opt} reveals some interesting results. First, the
duration of the line emission in the low-optical states is significantly shorter
than the continuum emission, that is, the changes in \ha\ occur on faster timescales than in the continuum. While the continuum emission may remain stable for
years during the optical minimum, the \ha\ line emission reaches a minimum and
begins to increase after just a few weeks. Second, the continuum emission reaches
its peak by about $\sim400$ days before the \ew\ does. Third, the source
exhibited stronger line and continuum emission  during the 2019 high state than
during the 2009 high state: \ew$=-13.6$ \AA\ and $R=13.74$ mag compared to
\ew$=-10$ \AA\ and $R=13.92$ mag.

\begin{figure} 
\resizebox{\hsize}{!}{\includegraphics{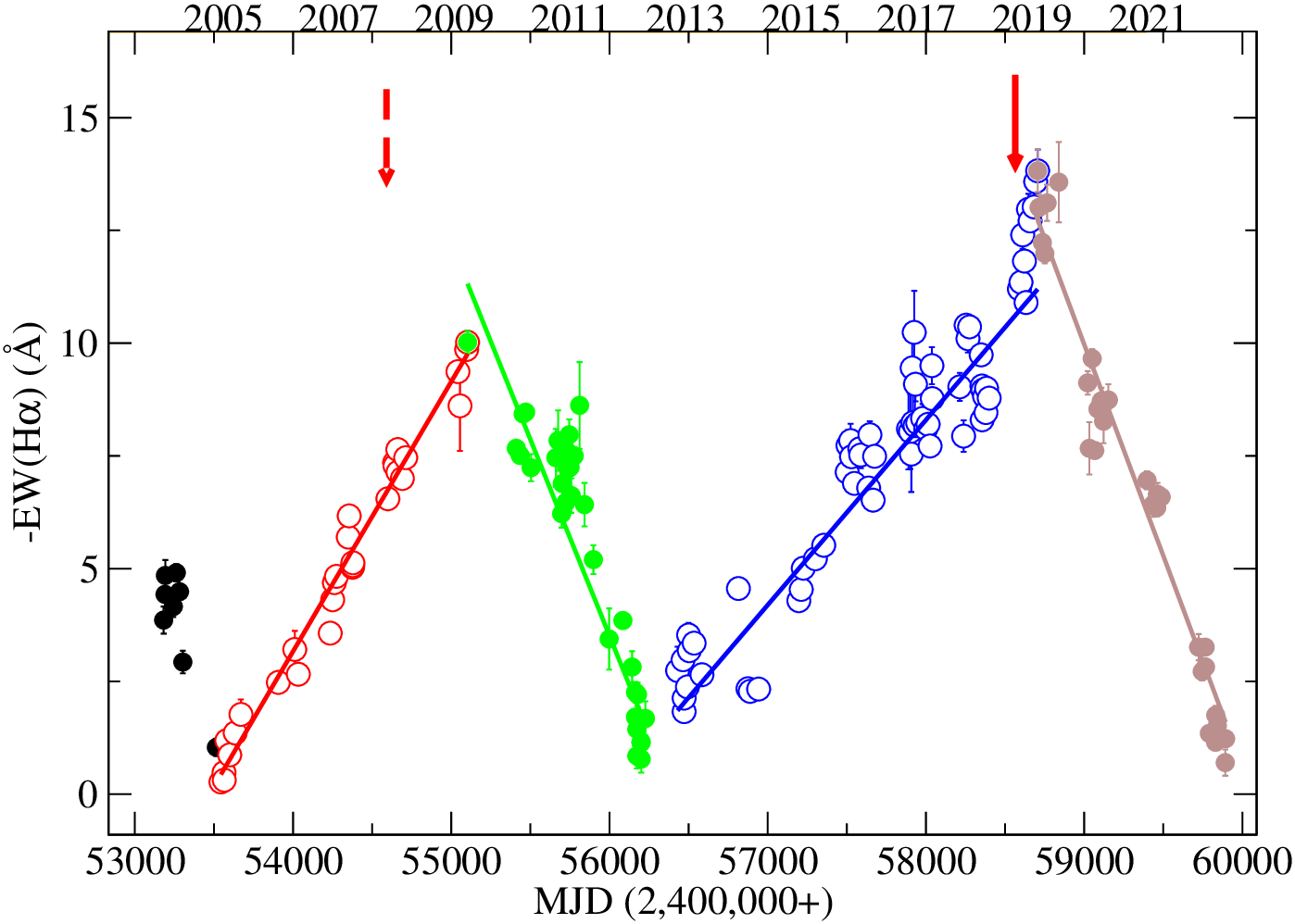}} 
\caption[]{Evolution of the \ha\ equivalent width. The circles represent
the data, and the solid lines are the best linear fit to the data points of the
corresponding interval. The red arrows mark the occurrence of an X-ray
outburst. The dash style simply denotes a weaker outburst.}
 \label{ewfit}
\end{figure}

The transition from low to high states and then back to low states occurred
smoothly over a period of several years. Overall, the rate of change in the \ew\ is
roughly similar during the two transitions from the low to the high states as
it is for the two declining phases. However, the changes in the optical
observables during the declining phases are in general
more abrupt that during the rising phases. To quantify this result, we performed a
linear fit to the \ew\ for the four different phases, as shown in
Fig.~\ref{ewfit}. The results are summarized in Table~\ref{linfit}. The slopes of
the fits are 1.5--2 times larger during the decay phase.

%-------------------------------Fig. 2-----------------------------------------------
%------------------------------------------------------------------------------

%-----------------------------------------------------------------------------------------
\begin{table*}
\caption{Results of the linear fits to the different parts of the long-term
evolution of the \ew. See Fig.~\ref{ewfit} to identify the various regions.}
\label{linfit}
\center
\begin{tabular}{lccccc}
\hline
\hline
Region  &MJD range      &slope                  &Ordinate       &Corr. Coef.  &$N_{\rm obs}$  \\
\hline
\hline
Rise 1 (red)    &53544--55103   &$0.00598\pm0.00006$    &$-320\pm3$     &0.99   &30   \\
Rise 2 (blue)   &56220--58751   &$0.00410\pm0.00004$    &$-230\pm2$     &0.95   &67   \\
Decay 2 (green) &55103--56220   &$-0.00870\pm0.00011$   &$491\pm6$      &-0.92  &33   \\
Decay 3 (brown) &58751--59894   &$-0.00938\pm0.00010$   &$560\pm5$      &-0.98  &31   \\
\hline
\end{tabular}
\end{table*}
%-----------------------------------------------------------------------------------------
%-------------------------------Fig. 3-----------------------------------------------
\begin{figure}
\begin{center}
\includegraphics[width=8cm]{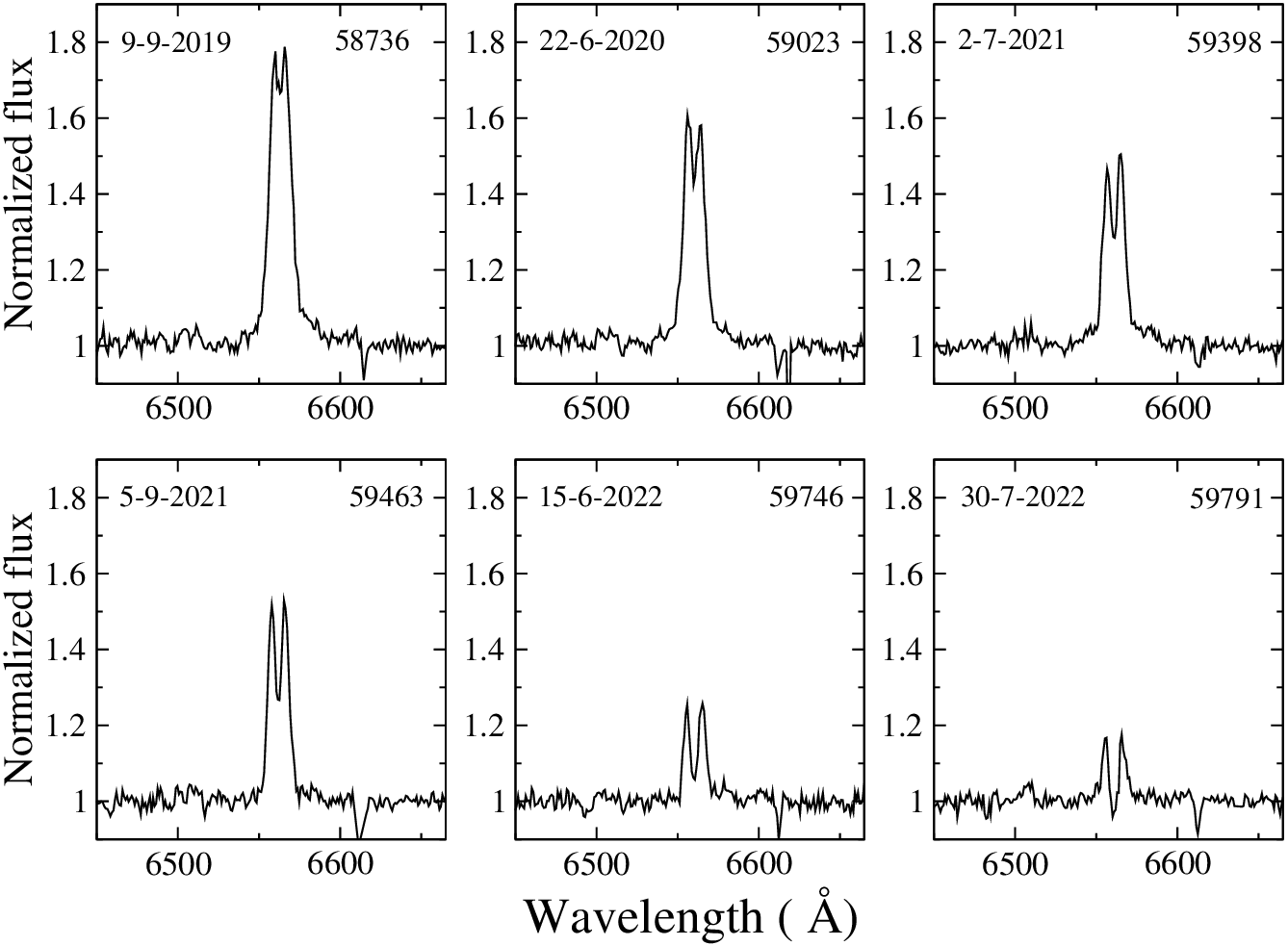} 
\caption[]{Representative profiles of the \ha\ line. }
\label{lineprof}
\end{center}
\end{figure}
%------------------------------------------------------------------------------
%--------------------------------Fig. 4----------------------------------------------
\begin{figure}
\begin{center}
\includegraphics[width=10cm]{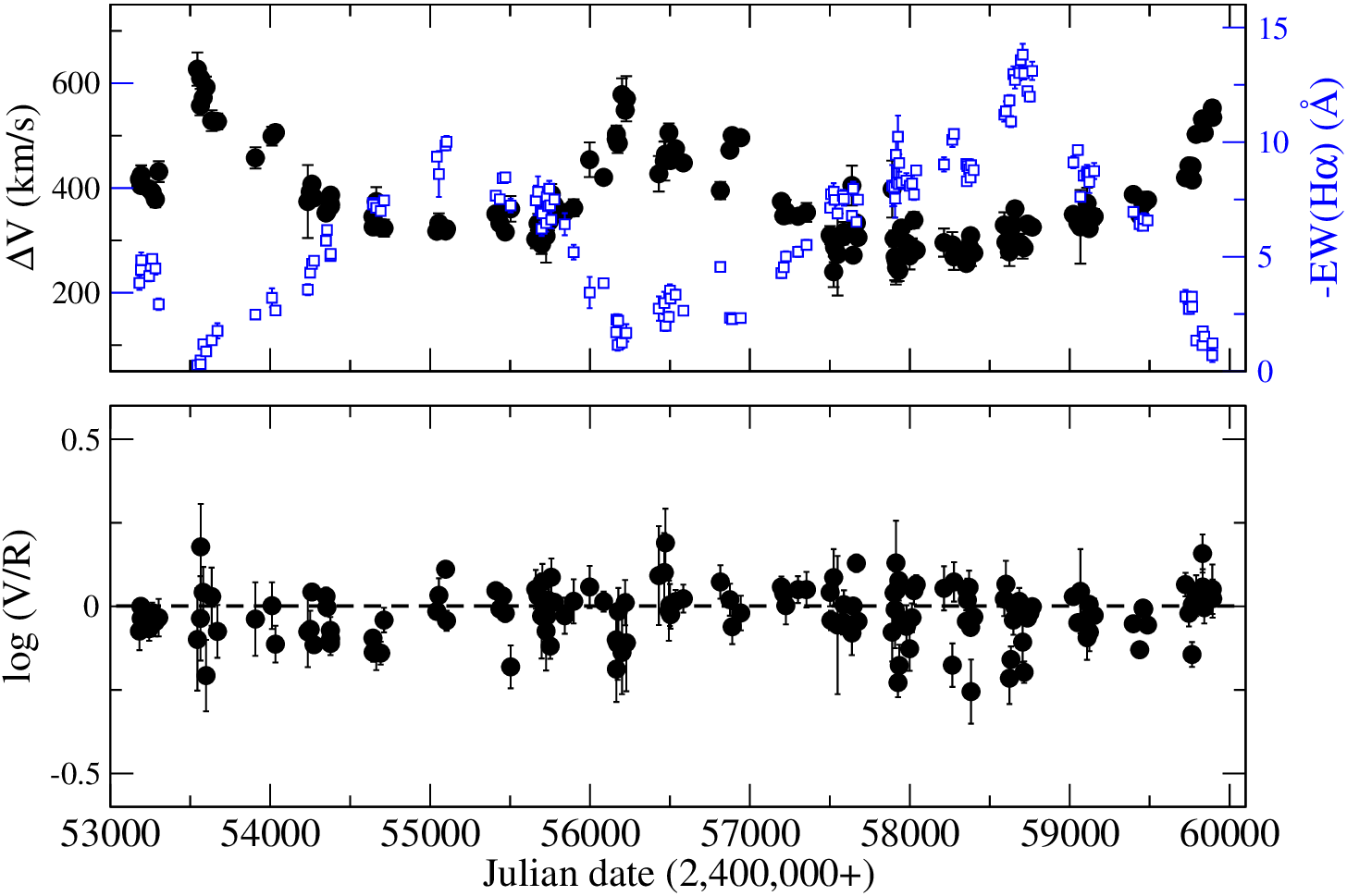} 
\caption[]{Evolution of the \ha\ line parameters. {\em Top}: Separation between the blue and 
red peaks (left axis) and the \ha\ EW (right axis). 
{\em Bottom}: Evolution of the $V/R$ ratio.  }
\label{linevol}
\end{center}
\end{figure}
%------------------------------------------------------------------------------

\subsection{\ha\ line profile variability}
\label{prof}

In contrast to what it is typically seen in other BeXBs, the shape of the \ha\ line
is rather symmetric.  While the strength of the line changes substantially over
time, as described in the previous section, the Be star in \gro\ always displays
a quite symmetric double-peak profile (Fig.~\ref{lineprof}). The ratio of the
peak intensities of the blue and red peaks is almost always $V/R\sim1$
(Fig.~\ref{linevol}, bottom panel), which it is rather unusual as most BeXBs
exhibit $V/R$ variability \citep{reig00,reig10b,yan16,alfonso17,zamanov20}. 

The
separation between the blue and red peaks $\Delta V$ also evolves with time
 and is anticorrelated with \ew\ (Fig.~\ref{linevol}, top panel). At the optical
high states, the peak separation approaches the spectral resolution of our data
($\sim200$ km s$^{-1}$), and  the two peaks merge into a single-peak profile in
some spectra. 

%We shall discuss the implications of these results in Sect~\ref{velaw}.

\subsection{Optical polarimetry}

The continuum polarization in Be stars is attributed to Thomson scattering of
unpolarized starlight in the disk
\citep{poeckert79,waters92,wood96,yudin01,halonen13a}.   In general, it is
difficult to compute the actual intrinsic polarization because most of the
observed polarization is expected to have an interstellar origin. For example,
using the relation between polarization and extinction $P_{\rm max,ISM} (\%) =
3.5 \times E(B-V)^{0.8}$ \citep{fosalba02}, we estimate that the maximum
contribution of the interstellar medium (ISM) to the measured optical
polarization toward \gro\ is $P_{\rm ISM}= 4.6$\%,  which is the observed
polarization, on average. However, the fact that  the polarization degree varies
by more than 1\% (from $\sim4$\% to $\sim5.2$\%) and that this variation
correlates with \ew\ (it increased as the \ew\ increased) supports the idea that
part of the observed polarization is intrinsic to the source and originates in
the disk. To estimate the contribution of the ISM, we observed three nearby
stars in the field of view of \gro. Table~\ref{polfsinf} lists the field stars
that we used, while Table~\ref{polfs} gives the polarimetric data for these
stars.

\begin{figure}
\begin{center}
\includegraphics[width=8cm]{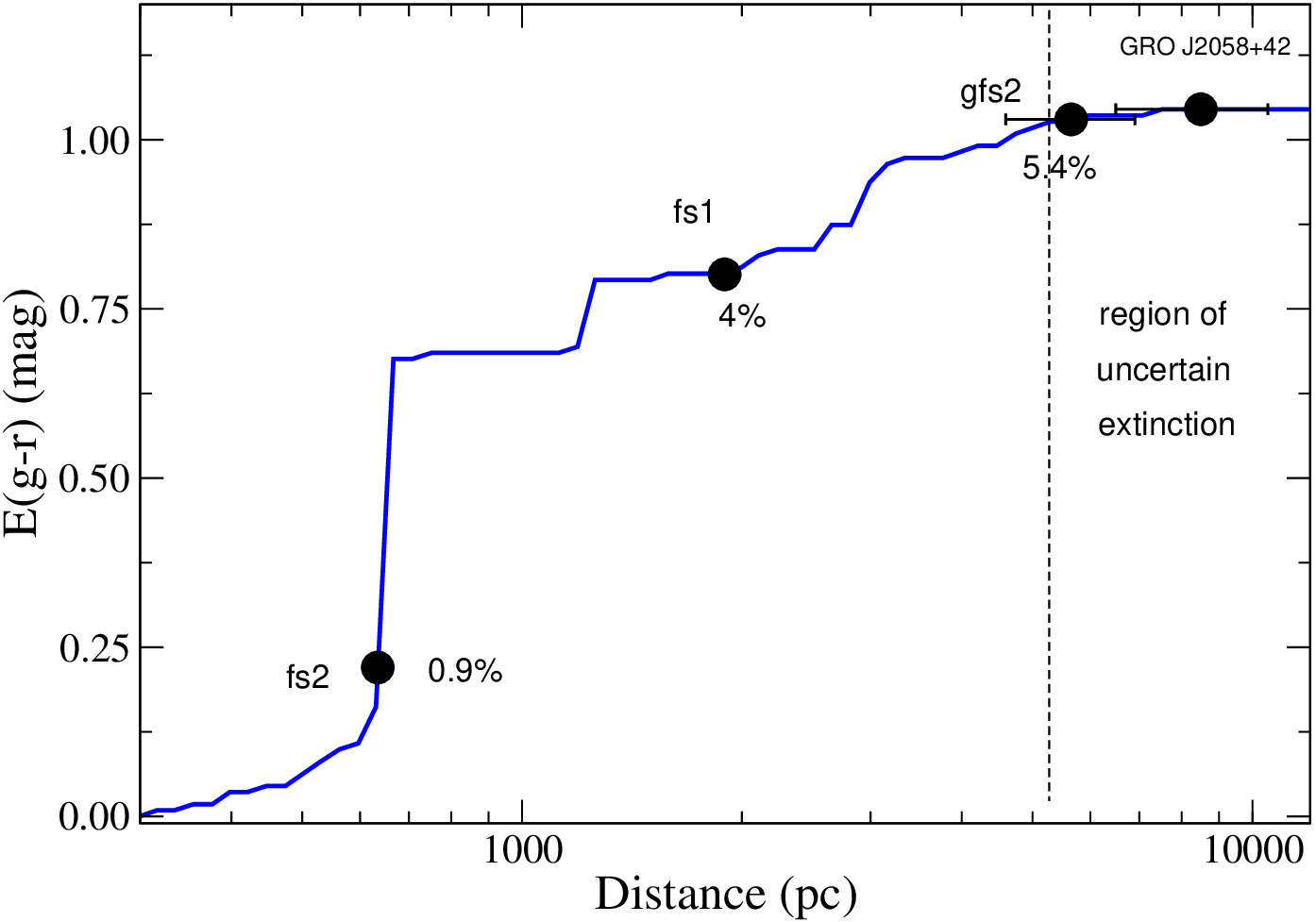} 
\caption[]{Extinction as a function of distance (data from \citet{green19}).
The dots mark the assumed location of the stars on the extinction curve,
according to their distance. Distances are from \citet{bailer-jones21}. We also
indicate the polarization degree of each star (see Table~\ref{polfs}).}
\label{extdist}
\end{center}
\end{figure}

Figure~\ref{extdist} shows the change in extinction with distance and the
location of the field stars and the source. The data were taken from the  dust
maps by \citet{green19}. The figure indicates that at least three different
molecular clouds are in between the observer and the BeXB along the line of
sight. Clearly, as the distance increases, so does the polarization. The field
star fs2 is too close to Earth to represent a reliable measurement of the ISM
polarization at the source distance.  On the other hand, gfs2 displays  stronger
polarization than the source, which might indicate that the star is
intrinsically polarized and hence again only a poor representation of the ISM
polarization. We note that at very short and long distances, the computation of
the extinction is very uncertain due to the lack of sufficient number of stars.
If we use fs1 to estimate the polarization from the ISM (by vector subtraction,
the $q$ and $u$ Stokes parameters of the source and the field star), then we
obtain that the intrinsic polarization varied from 0.3\% to 1.3\%. The lowest
value of the polarization degree corresponds to the 2013 and 2022 observations,
when the \ew\ displayed minimum values.  The level of intrinsic polarization
that we measure agrees with the expected  polarization in an axisymmetric
circumstellar disk predicted by single-scattering plus attenuation models, which
is typically $\simless 2$\% \citep{waters92}. It also agrees with the measured polarization
in a sample of classical Be stars \citep{yudin01}.

%------------------------------------
\begin{table}
\caption{Selected field stars in the field of view of \gro\ used to derived the
ISM polarization. }
\label{polfsinf}
\begin{center}
\begin{tabular}{@~l@~c@~ccc}
\noalign{\smallskip}    \hline\noalign{\smallskip}
ID      &RA             &DEC           &Distance        &{\it Gaia} ID  \\
        &               &              &(pc)            &\\
\noalign{\smallskip}    \hline\noalign{\smallskip}
fs1     &314.72292      &41.77444       &$1894^{+67}_{-48}$     &2065652877361883648  \\
fs2     &314.69167      &41.76639       &$642^{+7}_{-6}$        &2065653598916381568  \\
gfs2    &314.77043      &41.76491       &$5787^{+1771}_{-1095}$ &2065652671203453696  \\
%gfs3   &314.74485      &41.70786       &$5648^{+1257}_{-1052}$ &2065649540166770176  \\
\noalign{\smallskip}    \hline\noalign{\smallskip}
\end{tabular}
\end{center}
\tablefoot{Distance from \citet{bailer-jones21}}
\end{table}
%-----------------------------------

%------------------------------Fig 5------------------------------------------------
%------------------------------------------------------------------------------
%------------------------------------------------------------------------------
%\begin{figure}
%\begin{center}
%\includegraphics[width=8cm]{./figures/q-u_plane.eps} 
%\caption[]{Normalized observed Stokes parameters.  }
%\label{qu}
%\end{center}
%\end{figure}
%------------------------------------------------------------------------------

\subsection{Reddening and distance}

Owing to the presence of the circumstellar disk, the photometric magnitudes and
colors are affected by disk emission.  The overall effect is to make the star
appear redder, that is, to be seen under higher extinction than in the absence of
the disk. As a result, the estimate of the distance obtained from photometric
magnitudes that are thought to be contaminated by disk emission represents a lower
limit.  Some attempts have been made to quantify the contribution of the disk
to the photometric indices \citep{dachs88,fabregat90,riquelme12}. Here we take
advantage of the low optical states in \gro\ (Fig.~\ref{opt}). Because the
source of long-term optical variability is the disk, low optical states
correspond to instances in which the disk is almost absent. Therefore,  the low
optical states give us the opportunity to observe the underlying B star without
a major contribution from the disk. 

To estimate the distance through the distance-modulus relation, $V-M_V-A_V=5
\log(d)-5$, the amount of interstellar extinction $A_{\rm V}=R \times E(B-V)$ to
the source has to be determined.  The most direct method for estimating $E(B-V)$ is
to use the calibrated color of the star according to the spectral type. The
color excess is defined as $E(B-V)= (B-V)_{\rm obs}-(B-V)_0$, where $(B-V)_{\rm
obs}$ is the observed color and $(B-V)_0$ is the intrinsic color of the star.
The expected color for an O9.5-B0V star is $(B-V)_0=-0.29\pm0.02$. This value is
the average of the calibrations from
\citet{johnson66}, \citet{fitzgerald70}, \citet{gutierrez79}, \citet{wegner94},
and \citet{pecaut13}. The error is
the standard deviation of the five values.   We applied this method to the three
low optical states independently and obtained $8.6\pm0.9$ kpc, $7.8\pm0.8$ kpc,
and $8.0\pm0.9$ kpc for the 2005--2006, 2012--2013, and 2022--2023 low states.
We note, however, that the average \ha\ EW was lowest during
the 2005--2006 period at \ew=--0.9 \AA, compared to \ew=--3.3 \AA\ and \ew=--1.3 \AA\
in the other two low states, respectively. Hence we take $8.5\pm0.8$ kpc as the
final value for the distance. This value compares to $8.7^{+1.1}_{-0.9}$ kpc
from {\it Gaia} \citep{bailer-jones21}. The average color excess is estimated to be
$E(B-V)=1.42\pm0.02$.

%The error in the distance was obtained by propagation of the errors 0.02 mag in
%$V$ and $E(B-V)$ and 0.2 mag in $M_{\rm V}$. 

\section{Discussion}

BeXBs display variability on all timescales and at all wavelengths. The fastest
variability is detected in the X-rays (on the order of seconds) and corresponds
to the rotation of the neutron star, which manifests as pulsations. In the
optical band, the fastest timescales (on the order of hours or days) are
attributed to changes in the stellar photosphere (pulsation and rotation).
Long-term variations (on the order of months to years) are related to structural
changes in the disk. These long-term variations are the subject of the present
work.

\subsection{Disk formation and dissipation}

Be star disks are known to form and dissipate on timescales of years. The
disk emission increases with wavelength and becomes particularly significant in
the infrared band.
\gro\ went through two high (2008-2009 and 2018-2019) and three low optical
states (2004-2005, 2012-2013, and 2022-2023) during the course of our
observations. The transition from  low to high states and then back to low
states occurred smoothly over a period of several years. We attribute these
long-term changes to the formation and dissipation of the circumstellar disk. 
As expected in the disks of Be stars, the amplitude of the variability in the
continuum increases as the effective wavelength increases: the difference in
magnitudes from maximum to minimum is $\Delta B=0.32$, $\Delta V=0.46$, $\Delta
R=0.56$, and $\Delta I=0.70$.

%---------------------------------Fig. 6---------------------------------------------
\begin{figure}
\begin{center}
\includegraphics[width=8cm]{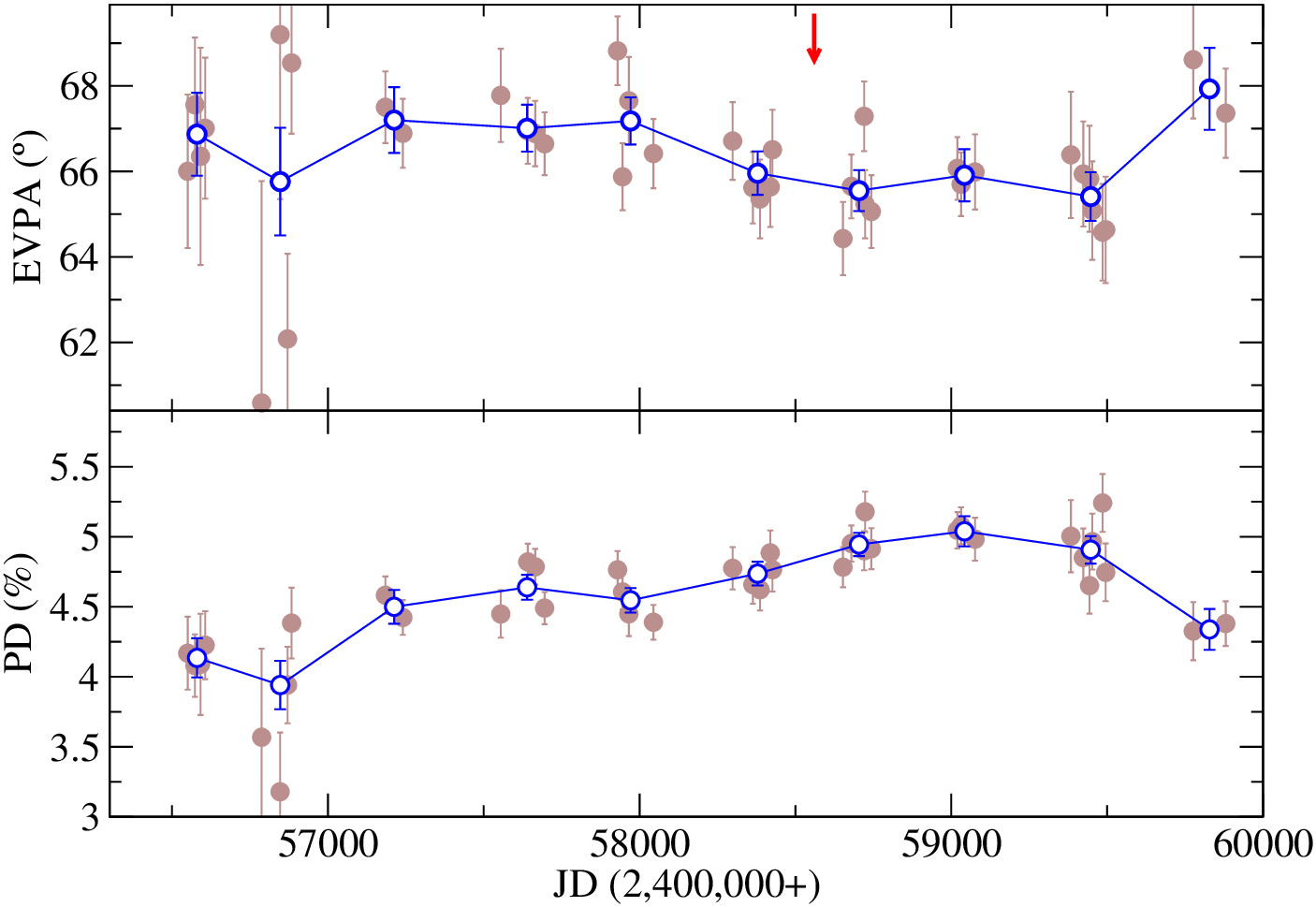} 
\caption[]{Polarization angle and polarization degree of \gro\ for the period
2013-2022. The red arrow marks the occurrence of a type II outburst. Filled brown circles denote all the observations, and the open blue circles
represent the weighted mean calculated using the observations over one year. }
\label{pol}
\end{center}
\end{figure}
%------------------------------------------------------------------------------

\subsubsection{Line and continuum emission}

In Sect.~\ref{res} we pointed out that the line emission variability timescales are faster than the continuum emission. The $BVRI$ magnitudes can remain
at a mean value with very small fluctuations during the low optical states. For
example, in the period MJD52800--54000, the mean and standard deviation of 14
measurements of the $V$ magnitude were 14.94 and 0.04 mag, respectively. For MJD
55800--56900 (six measurements), these values were 14.90 and 0.02 mag,
respectively. In contrast, \ew\ does not show a stable behavior like this and varies
more strongly. For the same periods, the average and standard deviation in
\ew\ was --1.8 and --0.8 \AA\ and --2.7 and --2.1 \AA, respectively. 

Another interesting result is the fact that the peak in the \ew\ occurred about
400 days after the peak in the photometric magnitudes. Although the
observational gaps cannot be ignored, the excellent spectral coverage  clearly
indicates that the delay between the \ew\  and $R$-band maxima is real. These
delays imply that the continuum and line emission are anticorrelated for a certain
period of time (i.e., the time elapsed between the two peaks). This result has
been reported in many other BeXBs: Swift\,J0243.6+6124 \citep{liu22},
RX\,J0440.9+4431 \citep{yan16}, 1A\,0535+26 \citep{clark99,yan12b}, 4U\,0115+63
\citep{reig07b}, and SAX\,J2103.5+4545 \citep{camero14}. Interestingly, the
X-ray outbursts occur close to the \ew\ maxima.

This delay can be explained as due to the different locations in the disk in which 
the line and continuum emission in Be stars originate. The outer parts of the
disk contribute most to the emission of the \ha\ line, whereas the origin of the
optical continuum resides in the inner parts. According to the viscous decretion
disk model \citep[see][for a review]{rivinius13a}, about 90--95\% of the
$V$-band flux comes from inside $1.8-2.5\,R_*$ \citep{haubois12}, while \ha\
emission reaches this percentage above $\simmore 6-10\, R_*$ \citep{carciofi11}.
Therefore, the fact that  the optical brightness has already started to decrease
by the time the \ew\ reached the maximum indicates that the dissipation of the
disk began from the inner parts of the disk. Disk dissipation occurs when the
mass ejection mechanism ceases. As the angular momentum supply stops, the inner
parts of the disk are reaccreted onto the star. However, viscosity still
transports angular momentum outward. Thus the outer parts of the disk continue
to expand, while the inner parts move inward \citep{haubois12,carciofi12}. Since
the flux in the $V$ (or $R$) band is mainly produced in the inner parts of the
disk, we expect the visual flux to start decreasing before the \ha\ line flux,
which mainly originates in the outer parts. 

The long-term trend of the polarization degree also supports a disk
formation and dissipation cycle. Polarization in Be stars results from the
scattering of stellar radiation by electrons in the circumstellar disk.
The degree of the polarization provides information about the number of
scatterers. Therefore, we should expect a significant decrease in the
polarization degree during low optical states, as is indeed observed
(Fig~\ref{opt}). Since these low optical states represent instances when the
disk almost vanished, we can assume that the polarization degree roughly
corresponds with that of the ISM. With this assumption, we estimate that the ISM
polarization from the observer to \gro\ is $\sim4$\%.

Strictly speaking, none of low optical states corresponded to the complete loss
of the disk because the \ha\ line did not revert fully into absorption (see
Fig.~\ref{lineprof}).  A normal (i.e., non-\ha\ emitting) B0V star is expected to
have \ew$=+3.5$ \AA\  \citep{jaschek87}, while the lowest values of \ew\ during
the low optical states were around $-[0.3-1]$ \AA. Nevertheless, the low optical
states do represent highly debilitated disk phases.

\subsubsection{Connection with X-rays}

As we mentioned above, the \ew\ during the 2018-2019 maximum reached higher
values than during the 2008-2009 high state. Since \ew\ provides a measure of
the size of the disk, we conclude that the Be star developed a larger disk
during the 2018-2019 event. The accretion of a large amount of matter onto the
neutron star led to the type II outburst observed by the {\it NuSTAR} and Astrosat space
missions. In principle, it might be argued that since the disk was smaller during the
2008-2009 optical peak, the X-ray emission would be lower, as observed.   On the
other hand, given the similarities and regularity of the low and high optical
states (Figs.~\ref{opt} and \ref{ewfit}), the absence of a giant X-ray outburst
prior to the 2008--2009 optical maximum is surprising. The {\it Swift}/BAT hard X-ray
transient monitor detected an increase in 15--50 keV flux by a factor of two in
less than a week. A pointed observation with {\it Swift}/XRT was made on 9 May 2008.
The 0.3--10 keV flux was estimated to be $1.5 \times 10^{-10}$ erg cm $^{-2}$
s$^{-1}$, which at a distance of 9 kpc corresponds to a luminosity of $L_x=1.4
\times 10^{36}$ \ergs\ \citep{krimm08}. Although this luminosity is one order of
magnitude lower than the 2019 event, we cannot be sure that it corresponded to
the peak of the X-ray event. The 2008--2009 is reminiscent of a type I outburst,
and the lack of data can be attributed to the weakness of the detection due to
the large distance to the source.

One possible explanation for the different luminosity of the two events could be
a different geometry in a misaligned and tilted disk. Current models that
explain X-ray outbursts in BeXBs invoke warped, tilted, and misaligned disks.
Because disks in BeXBs are truncated \citep{reig97a,reig16}, the only way for
large amount of material to be transferred to the neutron star is in highly
asymmetric configurations in which a misaligned disk becomes warped and eccentric.
Highly distorted disks result in enhanced mass accretion when the neutron star
moves across the warped part \citep{martin11,martin14a,okazaki13}. Evidence of these warped disks has been reported for 1A\,0535+262 \citep{moritani13} based
on a spectral line analysis and for 4U\,0115 \citep{reig18b} based on variations in the polarization
angle. Although we observe a small change in the polarization angle
during the 2019 X-ray outburst (Fig.~\ref{pol}), it is not very significant.
Likewise, the symmetry of the \ha\ line profile argues against a highly
distorted disk. Therefore, it remains to be seen how a type I outburst can lead
to the almost complete destruction of the circumstellar disk.

We also noted in Sect~\ref{res} that the dissipation of the disk is faster than
its formation. The rate of change of the \ha\ equivalent width is about
1.8 \AA/yr during disk formation and about 3.3 \AA/yr during
disk dissipation (see Fig.~\ref{ewfit} and Table~\ref{linfit}). This result
contrasts with what it is observed in classical Be stars, in which the
timescales for disk growth are shorter than the timescales for disk dissipation
\citep{haubois12}. A crucial difference is the presence of the neutron star in
BeXBs. As the disk expands and reaches periastron distance, a substantial amount
of matter will be accreted onto the neutron star. Therefore, the dissipation of
the disk in BeXBs may occur faster owing to the interaction between the neutron
star and the disk. 

%----------------------------------Fig. 7--------------------------------------------
\begin{figure}
\begin{center}
\includegraphics[width=8cm]{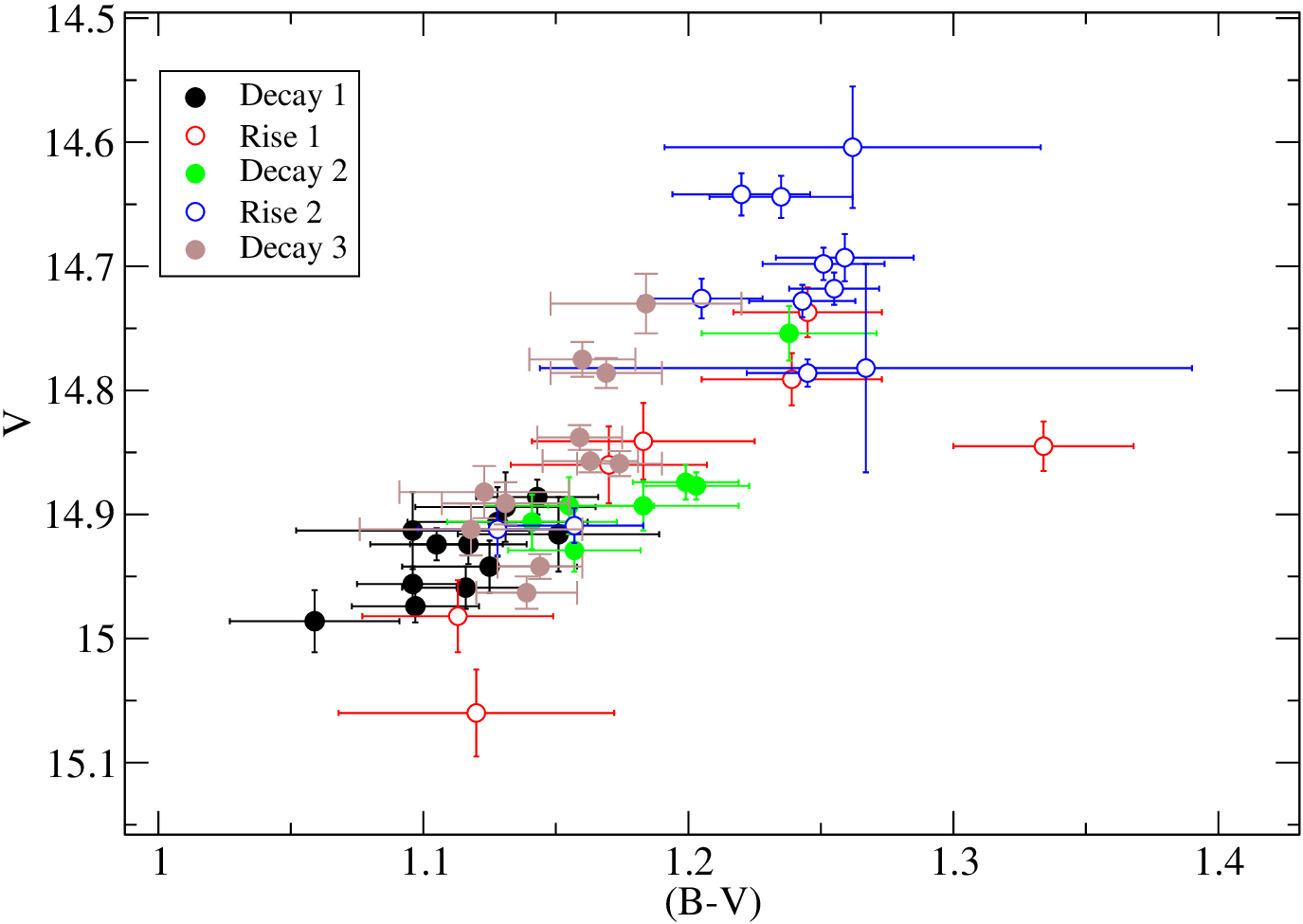} 
\caption[]{ ($(B-V)-V$ color magnitude diagram. Empty symbols correspond to
the disk formation phases, and filled symbols correspond to the dissipation
phases.}
\label{vbv}
\end{center}
\end{figure}
%------------------------------------------------------------------------------

\subsection{Inclination angle}
\label{ang}

We showed in Sect.~\ref{prof} that the source displays a double-peak profile even
during the high optical state (see Fig~\ref{lineprof}). This result suggests that we
see the circumstellar disk at an intermediate inclination angle. Typically,
single-peak profiles are associated with high inclination angles (pole-on
stars) and shell profiles with high inclination angles (edge-on stars;
\citealt{slettebak79,hanuschik95,rivinius06,silaj14,sigut22}). An example of a low
inclination system is V\,0332+54, whose inclination angle is estimated to be
$i<20^{\circ}$ \citep{negueruela99,zhang05}, and a BeXB that exhibits shell
profiles is IGR\,J21343+4738 \citep{reig14a}.

The evolution of the photometric magnitudes and colors can also be used to
constrain the inclination angle. Figure~\ref{vbv} shows the color-magnitude
diagram of \gro. Empty symbols correspond to the disk formation phases, and
filled symbols correspond to the dissipation phases. The positive correlation
between the color index $(B-V)$ and the $V$ magnitude implies that the system is
viewed at an intermediate inclination angle, where the disk emission to the
optical colors and magnitudes contributes significantly. As the disk grows, the
system brightens ($V$ decreases), while the overall (disk plus star) emission
becomes redder ($(B-V)$ increases).  At very high inclination angles (equator-on
stars), the inner parts of the Be envelope partly block the stellar photosphere,
while the small projected area of the disk on the sky keeps the disk emission to
a minimum. Thus, stars viewed at very high inclination angles ($i > i_{\rm
crit}$) would show an inverse correlation \citep{harmanec83,harmanec00}.   The
value of the critical inclination angle is not known, but a rough estimate based
on available data suggests $i_{\rm crit} \sim 60-70^{\circ}$
\citep{hanuschik96a,sigut13}.
In summary, the correlation between the photometric colors and magnitudes and
the stable double peak profile of the \ha\ line in \gro\ suggest an inclination
angle in the range $30^{\circ}-60^{\circ}$.

%-----------------------------------Fig 8-------------------------------------------
\begin{figure}
\begin{center}
\includegraphics[width=8cm]{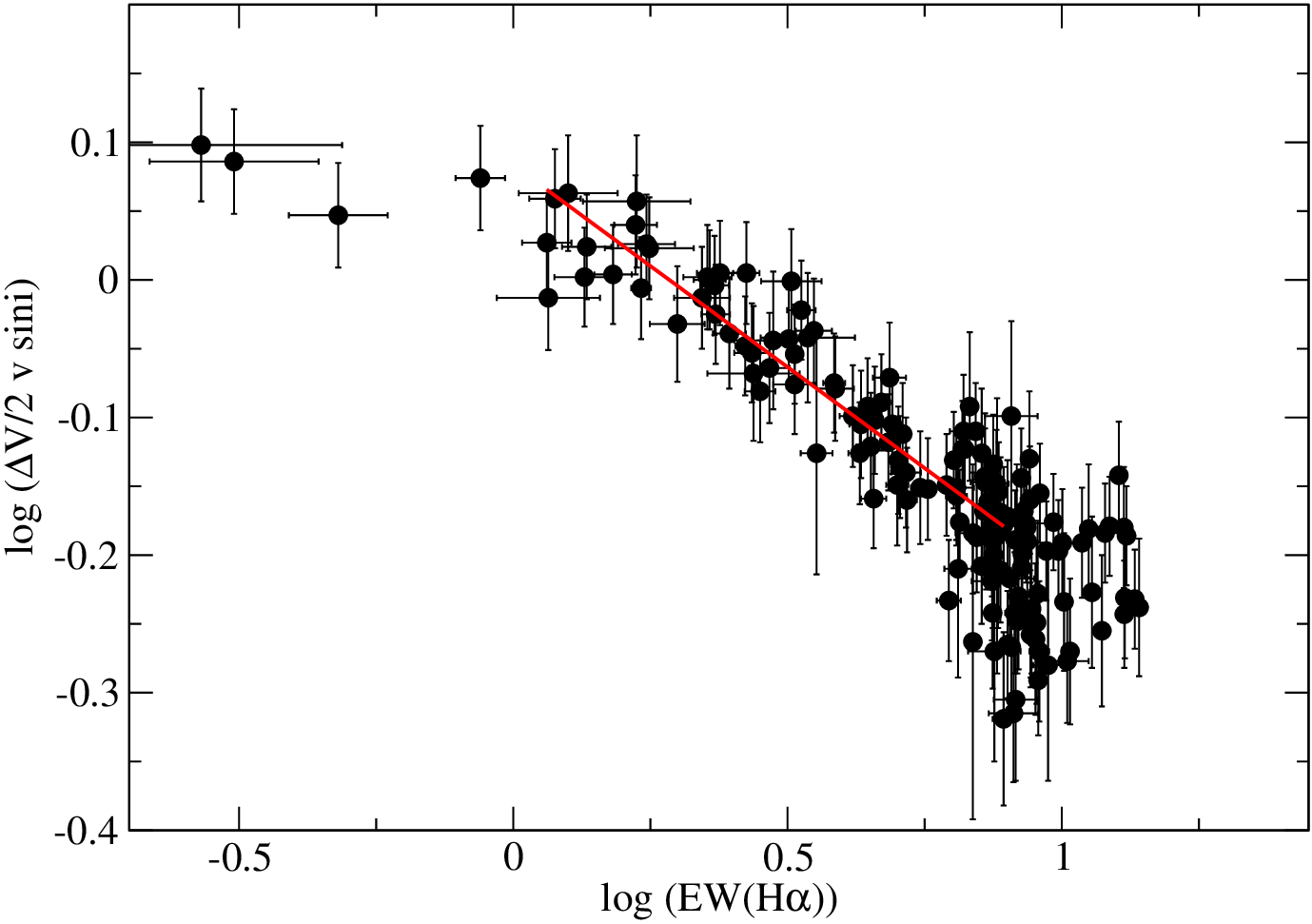} 
\caption[]{Peak separation as a function of \ew.  }
\label{deltav-ew}
\end{center}
\end{figure}
%------------------------------------------------------------------------------

\subsection{Disk size and kinematics}
\label{velaw}

%Emission lines represent the defining property of Be stars. They are known to
%provide fundamental information about the physical properties of the
%circumstellar disk. Among all lines, \ha\ stands up as the most important line
%because it is the strongest of the hydrogen hydrogen series and easily
%detectable in optical spectra, even at low spectral resolution. Among the most
%important physical disk parameters that can be investigated from the strength and
%profile of the \ha\ line are its size, geometry, kinematics, and density
%structure.

%The strength of the line provide information about the extension of the disk
%\citep{tycner05,grundstrom06}

It was already proposed by \citet{struve31} that disks in Be stars are supported by rotation. The main evidence for this comes from the
correlation between the projected rotational velocity and the width and shape of
the emission line \citep{andrillat83,dachs86b,hanuschik88,hanuschik89,dachs92}.

In Sect~\ref{prof} we showed that the peak
separation $\Delta V$ and \ew\ are anticorrelated. This is the expected behavior in rotation-supported disks, in which the rotational velocity of the gas particles in the disk
varies as the inverse of the radius, $v_{\phi}\propto r^{-j}$  (e.g., for
Keplerian disks $j=1/2$).  In these disks, the peak separation roughly
corresponds to twice the projected rotational velocity of the particles in the
disk $\Delta V\approx 2 v_{\phi} \sin i$. As the radius of the disk increases
(i.e., as \ew\ increases), the rotational velocity decreases, and so does the peak
separation.  The index $j$ can be estimated from the relation between peak
separation and EW \citep{hanuschik88},

\begin{equation}
log(\Delta V/2 \sin i) = a \times \log(EW(H\alpha))+b
,\end{equation}

\noindent where the slope $a$ is related to the $j$ index as $a=j/2$. Figure \ref{deltav-ew} shows
this relation for \gro. When we restrict the fitting range to EWs between $1-8$ \AA, then we find $j=0.58\pm0.04$, which is not far from the
expected value of 0.5 for a Keplerian disk. Lower and higher values of the \ew\ were removed from the fit because below $\log(EW(H\alpha))\sim
0$ the disk is too small and has probably not reached a stable configuration,
while above 8 \AA, the $\Delta V$ saturates, which might be a spectral
resolution effect.

\begin{figure}
\begin{center}
\includegraphics[width=8cm]{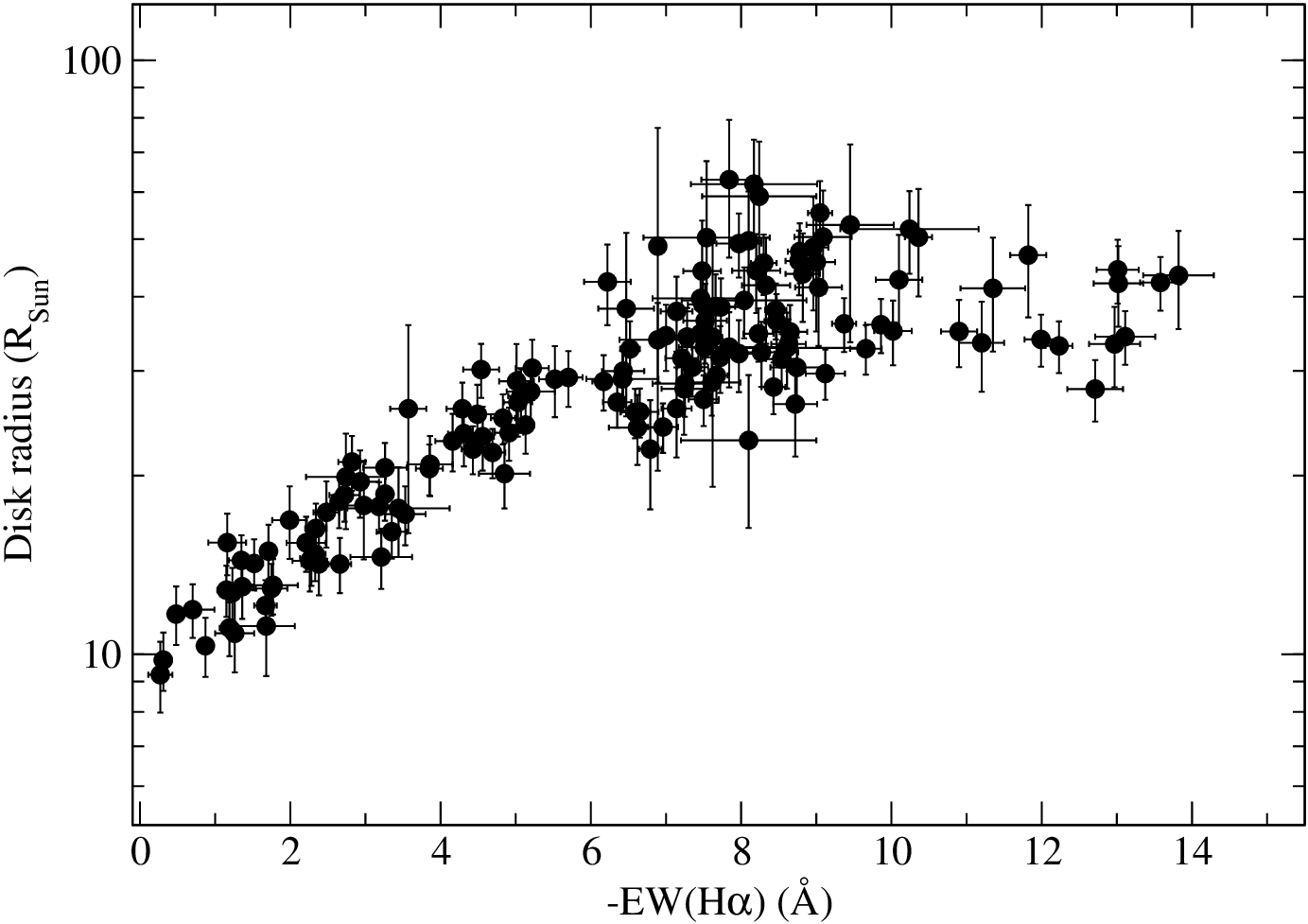} 
\caption[]{Disk radius as a function of \ew. }
\label{radew}
\end{center}
\end{figure}

After showing that the particles in the disk of \gro\ follow a Keplerian
law, we can estimate the disk size. The disk velocity adopts the form 
\citep[see. e.g.,][]{huang72} 

\begin{equation}
\label{vphi}
v_\phi=v_0\left(\frac{r}{R_*}\right)^{-1/2}
,\end{equation}

\noindent where $R_*$ is the equatorial stellar radius, and $v_0$ is a initial
value of the disk velocity close to the stellar surface. In the limiting case,
$v_0$ would be the stellar rotational velocity. For simplicity, we
assumed $v_0=v_*$. By inverting eq.~(\ref{vphi}) and since 
$v_{\phi} \approx (\Delta V/2 \sin i),$

\begin{equation}
\frac{R_d}{R_*} = \left(\frac{2 v_* \sin i}{\Delta V}\right)^2 
,\end{equation}

\noindent where $r=R_d$ is the radius of the \ha\ emission region.
Figure~\ref{radew} shows the size of the circumstellar disk as a function of
time and \ew. As mentioned above, the leveling off at $\sim 50\,\rsun$ marks the
minimum distance between the peaks that our spectrograph is capable of
discerning.

The Be star projected rotational velocity in \gro\ has been estimated by
\citet{kiziloglu07b} to be in the range $240-310$ km s$^{-1}$. Here we assumed $v_*
\sin i = 275$ km s$^{-1}$ , which corresponds to half the maximum peak separation.
We note, however, that \ha\ double-peak separation exceeding $2 v_* \sin i$ has
been measured in Be stars during phases of very weak \ha\ emission
\citep{dachs92}.

The largest disk radius is estimated to be $R_{\rm disk,max}\sim5-6 \, R_*$ or
$60\, \rsun$, assuming a stellar radius of an O9.5-B0e IV-V star of $10\, \rsun$
\citep{martins05}.  From Kepler's third law, we can estimate the orbital
separation $a\sim 160 \, \rsun$ for $P_{\rm orb}=55$ days and typical values of
the mass $M_*=18\, \msun$. If we now assume that in order to produce an X-ray
outburst, the neutron star must interact with the disk, that is, the disk radius
must be similar to the periastron distance ($a(1-e)$), then the eccentricity of
the system should be $e\sim 0.6$.

%---------------------------------Fig. 9---------------------------------------------
%------------------------------------------------------------------------------

\section{Conclusion}

We have performed the most detailed study of the long-term optical variability
of the BeXB \gro.  The system displays correlated variability on timescales of
years in all the optical observables. The source goes through optical high and
low states. The high optical state is characterized by large \ew\, bright
continuum emission, and significant polarization. We identify this state with a
well-developed decretion disk. The smooth increases and decreases of the optical
emission are interpreted as phases of formation and dissipation of the
circumstellar disk. The optical maxima coincide with the occurrence of X-ray
outbursts, after which all optical indicators decreased, leading to the
dissipation of the disk.  However, a disk-loss episode never occurred.  The
almost permanent double-peak line profile indicates that the systems is seen at
an intermediate inclination angle, whereas the absence of $V/R$ variability is
attributed to a relatively small and stable disk. We did not find evidence for a
warped disk during the 2019 type II X-ray outburst.

\begin{acknowledgements}

Skinakas Observatory is run by the University of Crete and the Foundation for
Research and Technology-Hellas.   The Liverpool Telescope is owned and operated
by the Astrophysics Research Institute of Liverpool John Moores University. The
Starlink software is currently supported by the East Asian Observatory. IRAF is
distributed by the National Optical Astronomy Observatories, which is operated
by the Association of Universities for Research in Astronomy, Inc. (AURA) under
cooperative agreement with the National Science Foundation. This research has
made use of the SIMBAD database, operated at CDS, Strasbourg, France and of
NASA's Astrophysics Data System operated by the Smithsonian Astrophysical
Observatory.

\end{acknowledgements}

\bibliographystyle{aa}
\bibliography{../../../artBex_bib}

\clearpage
\onecolumn

\begin{appendix}
\section{Results of the observations}
\label{app}

In this section we present the results of our optical data analysis.

%------------------------------------
\begin{table*}
\caption{Photometric observations of the optical counterpart to \gro.}
\label{photres}
\begin{center}
\begin{tabular}{cccccccccc}
\noalign{\smallskip}    \hline\noalign{\smallskip}
Date &  JD (2,400,000+)&   $B$  &errB   &$V$   &errV   &$R$  &errR &$I$ &errI  \\
\noalign{\smallskip}    \hline\noalign{\smallskip}
07-06-2003 & 52798.507 & 16.009 & 0.031 & 14.913 & 0.031 & 14.219 & 0.027 & 0.000 & 0.000   \\
08-06-2003 & 52799.439 & 16.067 & 0.025 & 14.942 & 0.021 & 14.241 & 0.022 & 0.000 & 0.000   \\
24-08-2003 & 52876.381 & 16.034 & 0.020 & 14.906 & 0.028 & 14.259 & 0.025 & 13.499 & 0.034   \\
24-08-2003 & 52876.381 & 16.025 & 0.020 & 14.894 & 0.028 & 14.247 & 0.025 & 13.479 & 0.034   \\
05-07-2004 & 53192.458 & 16.029 & 0.018 & 14.886 & 0.014 & 14.156 & 0.017 & 13.350 & 0.024   \\
27-07-2004 & 53214.509 & 16.067 & 0.024 & 14.916 & 0.030 & 14.193 & 0.027 & 13.411 & 0.041   \\
24-08-2004 & 53242.434 & 16.052 & 0.015 & 14.956 & 0.014 & 14.244 & 0.012 & 13.456 & 0.017   \\
14-09-2004 & 53263.350 & 16.041 & 0.015 & 14.924 & 0.016 & 14.220 & 0.014 & 13.468 & 0.026   \\
01-10-2004 & 53280.301 & 16.075 & 0.017 & 14.959 & 0.017 & 14.238 & 0.018 & 13.474 & 0.019   \\
26-06-2005 & 53548.515 & 16.029 & 0.021 & 14.924 & 0.013 & 14.224 & 0.018 & 13.463 & 0.044   \\
27-07-2005 & 53579.475 & 16.071 & 0.020 & 14.974 & 0.013 & 14.275 & 0.014 & 0.000 & 0.000   \\
20-08-2005 & 53603.383 & 16.045 & 0.020 & 14.986 & 0.025 & 14.282 & 0.024 & 0.000 & 0.000   \\
17-08-2006 & 53965.390 & 16.180 & 0.038 & 15.060 & 0.035 & 14.300 & 0.051 & 0.000 & 0.000   \\
19-08-2006 & 53967.436 & 16.095 & 0.021 & 14.982 & 0.029 & 14.291 & 0.034 & 0.000 & 0.000   \\
25-06-2007 & 54277.509 & 16.030 & 0.020 & 14.860 & 0.031 & 14.080 & 0.025 & 0.000 & 0.029   \\
16-07-2007 & 54298.476 & 16.179 & 0.028 & 14.845 & 0.020 & 14.081 & 0.020 & 13.245 & 0.000   \\
01-09-2007 & 54345.481 & 16.030 & 0.027 & 14.791 & 0.021 & 14.013 & 0.025 & 13.187 & 0.041   \\
02-09-2007 & 54346.438 & 16.024 & 0.028 & 14.841 & 0.031 & 14.042 & 0.020 & 13.203 & 0.021   \\
03-09-2007 & 54347.562 & 16.608 & 0.013 & 15.138 & 0.013 & 14.430 & 0.017 & 13.637 & 0.023   \\
05-08-2008 & 54684.414 & 15.982 & 0.019 & 14.737 & 0.020 & 13.922 & 0.020 & 13.062 & 0.023   \\
29-06-2009 & 55012.457 & 15.992 & 0.025 & 14.754 & 0.022 & 13.943 & 0.018 & 13.095 & 0.018   \\
26-07-2010 & 55404.420 & 16.076 & 0.030 & 14.893 & 0.020 & 14.164 & 0.025 & 13.368 & 0.028   \\
27-07-2010 & 55405.537 & 16.080 & 0.017 & 14.877 & 0.011 & 14.124 & 0.013 & 13.305 & 0.014   \\
02-11-2010 & 55503.310 & 16.270 & 0.019 & 15.356 & 0.020 & 14.432 & 0.018 & 13.490 & 0.021   \\
26-08-2011 & 55800.369 & 16.073 & 0.014 & 14.874 & 0.014 & 14.146 & 0.015 & 13.364 & 0.025   \\
09-09-2011 & 55814.256 & 16.048 & 0.022 & 14.893 & 0.023 & 14.194 & 0.015 & 13.433 & 0.032   \\
29-07-2013 & 56503.413 & 16.086 & 0.019 & 14.929 & 0.017 & 14.243 & 0.019 & 13.498 & 0.022   \\
29-08-2013 & 56534.3870 & 16.047 & 0.023 & 14.906 & 0.022 & 14.193 & 0.023 & 13.419 & 0.039   \\
20-08-2014 & 56890.4088 & 16.040 & 0.021 & 14.912 & 0.021 & 14.206 & 0.020 & 13.465 & 0.027   \\
14-09-2014 & 56915.3655 & 16.066 & 0.022 & 14.909 & 0.014 & 14.183 & 0.017 & 13.410 & 0.027   \\
22-07-2015 & 57226.4616 & 16.031 & 0.020 & 14.786 & 0.011 & 14.009 & 0.010 & 13.155 & 0.012   \\
18-11-2015 & 57345.2298 & 15.971 & 0.015 & 14.728 & 0.013 & 13.934 & 0.015 & 13.085 & 0.015   \\
06-06-2016 & 57546.4575 & 15.931 & 0.017 & 14.726 & 0.016 & 13.897 & 0.017 & 13.033 & 0.029   \\
22-06-2016 & 57562.5243 & 15.949 & 0.019 & 14.698 & 0.013 & 13.890 & 0.012 & 13.012 & 0.029   \\
08-09-2016 & 57640.2994 & 16.049 & 0.090 & 14.782 & 0.084 & 13.970 & 0.086 & 13.128 & 0.093   \\
06-10-2016 & 57668.2905 & 15.973 & 0.011 & 14.718 & 0.013 & 13.883 & 0.012 & 12.993 & 0.013   \\
03-11-2016 & 57696.3765 & 15.952 & 0.018 & 14.693 & 0.019 & 13.861 & 0.020 & 13.008 & 0.024   \\
28-08-2017 & 57994.3935 & 15.862 & 0.020 & 14.642 & 0.017 & 13.835 & 0.018 & 12.979 & 0.025   \\
13-06-2018 & 58283.4055 & 15.866 & 0.051 & 14.604 & 0.049 & 13.736 & 0.056 & 12.865 & 0.070   \\
31-07-2018 & 58331.4022 & 15.879 & 0.021 & 14.644 & 0.017 & 13.813 & 0.020 & 13.002 & 0.033   \\
13-07-2019 & 58678.3712 & 15.914 & 0.027 & 14.730 & 0.024 & 13.915 & 0.022 & 13.062 & 0.036   \\
30-07-2019 & 58695.3461 & 15.935 & 0.014 & 14.775 & 0.014 & 13.964 & 0.011 & 13.143 & 0.019   \\
11-09-2019 & 58738.3514 & 15.955 & 0.017 & 14.786 & 0.012 & 14.014 & 0.014 & 13.165 & 0.022   \\
24-06-2020 & 59025.4564 & 16.020 & 0.016 & 14.857 & 0.009 & 14.107 & 0.010 & 13.319 & 0.013   \\
19-07-2020 & 59050.3951 & 15.997 & 0.013 & 14.838 & 0.010 & 14.094 & 0.012 & 13.311 & 0.021   \\
19-08-2020 & 59081.3532 & 16.033 & 0.012 & 14.859 & 0.010 & 14.128 & 0.008 & 13.326 & 0.013   \\
04-07-2021 & 59400.3893 & 16.022 & 0.017 & 14.891 & 0.017 & 14.184 & 0.018 & 13.441 & 0.022   \\
04-09-2021 & 59462.3711 & 16.005 & 0.024 & 14.882 & 0.021 & 14.174 & 0.022 & 13.423 & 0.031   \\
10-10-2021 & 59498.3211 & 16.030 & 0.036 & 14.912 & 0.021 & 14.231 & 0.022 & 13.466 & 0.026   \\
29-07-2022 & 59790.3894 & 16.086 & 0.013 & 14.942 & 0.010 & 14.271 & 0.015 & 13.547 & 0.025   \\
17-08-2022 & 59809.3586 & 16.102 & 0.014 & 14.963 & 0.013 & 14.267 & 0.019 & 13.563 & 0.020   \\
20-09-2022 & 59843.3477 & 16.074 & 0.015 & 14.944 & 0.013 & 14.271 & 0.013 & 13.536 & 0.019   \\
\noalign{\smallskip}    \hline\noalign{\smallskip}
\end{tabular}
\end{center}
\end{table*}
%-----------------------------------

%------------------------------------
\begin{table*}
\caption{Polarimetric observations of the optical counterpart to \gro}
\label{polres}
\begin{center}
\begin{tabular}{cccccccccc}
\noalign{\smallskip}    \hline\noalign{\smallskip}
Date &  JD (2,400,000+) &PD (\%)        &err    &EPVA (deg.)    &err    &$q$    &err    &$u$    &err  \\
\noalign{\smallskip}    \hline\noalign{\smallskip}
14-09-2013   &56550.4068   & 4.17   &0.26  &66.00   &1.80    &-0.0279     &0.0028     &0.0310      &0.0028 \\ 
07-10-2013   &56573.4279   & 4.08   &0.22  &67.56   &1.57    &-0.0289     &0.0027     &0.0288      &0.0022 \\ 
25-10-2013   &56591.3649   & 4.09   &0.36  &66.35   &2.54    &-0.0277     &0.0039     &0.0301      &0.0036 \\ 
09-11-2013   &56606.2704   & 4.23   &0.24  &67.01   &1.65    &-0.0294     &0.0027     &0.0304      &0.0026 \\ 
10-05-2014   &56787.6089   & 3.57   &0.63  &60.58   &5.19    &-0.0185     &0.0077     &0.0305      &0.0057 \\ 
08-07-2014   &56846.5993   & 3.18   &0.42  &69.20   &3.85    &-0.0238     &0.0042     &0.0211      &0.0044 \\ 
01-08-2014   &56870.5431   & 3.94   &0.27  &62.08   &1.99    &-0.0221     &0.0028     &0.0326      &0.0029 \\ 
14-08-2014   &56883.5078   & 4.38   &0.25  &68.54   &1.66    &-0.0321     &0.0028     &0.0299      &0.0027 \\ 
11-06-2015   &57184.5220   & 4.58   &0.13  &67.50   &0.84    &-0.0324     &0.0018     &0.0324      &0.0018 \\ 
06-08-2015   &57240.5489   & 4.42   &0.12  &66.89   &0.81    &-0.0306     &0.0018     &0.0319      &0.0016 \\ 
15-06-2016   &57554.5689   & 4.45   &0.17  &67.78   &1.09    &-0.0318     &0.0021     &0.0312      &0.0019 \\ 
09-09-2016   &57641.4888   & 4.82   &0.13  &66.95   &0.77    &-0.0334     &0.0018     &0.0347      &0.0018 \\ 
03-10-2016   &57665.2876   & 4.79   &0.13  &66.89   &0.76    &-0.0331     &0.0018     &0.0346      &0.0017 \\ 
02-11-2016   &57695.2729   & 4.49   &0.12  &66.65   &0.73    &-0.0308     &0.0016     &0.0327      &0.0016 \\ 
24-06-2017   &57929.4772   & 4.76   &0.13  &68.82   &0.80    &-0.0352     &0.0018     &0.0321      &0.0017 \\ 
10-07-2017   &57945.4809   & 4.61   &0.13  &65.87   &0.79    &-0.0307     &0.0018     &0.0344      &0.0016 \\ 
31-07-2017   &57965.5018   & 4.45   &0.16  &67.65   &1.03    &-0.0316     &0.0022     &0.0313      &0.0016 \\ 
17-10-2017   &58044.2878   & 4.39   &0.12  &66.42   &0.81    &-0.0298     &0.0017     &0.0322      &0.0016 \\ 
29-06-2018   &58298.5334   & 4.78   &0.15  &66.71   &0.91    &-0.0328     &0.0021     &0.0347      &0.0018 \\ 
01-09-2018   &58363.3993   & 4.66   &0.14  &65.62   &0.84    &-0.0307     &0.0020     &0.0350      &0.0016 \\ 
24-09-2018   &58386.4209   & 4.62   &0.15  &65.35   &0.92    &-0.0301     &0.0021     &0.0350      &0.0017 \\ 
27-10-2018   &58419.2449   & 4.89   &0.16  &65.64   &0.94    &-0.0322     &0.0022     &0.0367      &0.0018 \\ 
03-11-2018   &58426.2604   & 4.76   &0.16  &66.51   &0.93    &-0.0325     &0.0021     &0.0348      &0.0018 \\ 
17-06-2019   &58652.4635   & 4.78   &0.14  &64.43   &0.86    &-0.0300     &0.0020     &0.0372      &0.0018 \\ 
14-07-2019   &58679.4369   & 4.95   &0.13  &65.65   &0.74    &-0.0327     &0.0018     &0.0372      &0.0018 \\ 
24-08-2019   &58720.4205   & 4.90   &0.14  &67.29   &0.82    &-0.0344     &0.0019     &0.0349      &0.0018 \\ 
27-08-2019   &58723.3256   & 5.18   &0.14  &65.23   &0.80    &-0.0336     &0.0020     &0.0394      &0.0019 \\ 
16-09-2019   &58743.3588   & 4.91   &0.15  &65.06   &0.85    &-0.0317     &0.0020     &0.0376      &0.0018 \\ 
19-06-2020   &59019.5386   & 5.05   &0.13  &66.07   &0.73    &-0.0339     &0.0019     &0.0374      &0.0018 \\ 
01-07-2020   &59031.5139   & 5.08   &0.13  &65.70   &0.74    &-0.0336     &0.0019     &0.0381      &0.0018 \\ 
14-08-2020   &59076.2758   & 4.98   &0.15  &65.99   &0.88    &-0.0333     &0.0020     &0.0370      &0.0019 \\ 
18-06-2021   &59383.5611   & 5.00   &0.26  &66.39   &1.48    &-0.0340     &0.0026     &0.0367      &0.0031 \\ 
27-07-2021   &59423.4226   & 4.85   &0.21  &65.94   &1.23    &-0.0324     &0.0027     &0.0361      &0.0021 \\ 
16-08-2021   &59443.3635   & 4.65   &0.20  &65.83   &1.24    &-0.0309     &0.0026     &0.0348      &0.0020 \\ 
25-08-2021   &59452.3016   & 4.97   &0.20  &65.08   &1.15    &-0.0320     &0.0027     &0.0379      &0.0021 \\ 
27-09-2021   &59485.4348   & 5.24   &0.21  &64.58   &1.13    &-0.0331     &0.0028     &0.0406      &0.0022 \\ 
07-10-2021   &59495.3763   & 4.75   &0.21  &64.63   &1.24    &-0.0300     &0.0027     &0.0367      &0.0021 \\ 
15-07-2022   &59776.4256   & 4.12   &0.11  &66.63   &0.76    &-0.0282     &0.0015     &0.0300      &0.0015 \\ 
\noalign{\smallskip}    \hline\noalign{\smallskip}
\end{tabular}
\end{center}
\end{table*}
%-----------------------------------

%------------------------------------
\begin{table*}
\caption{Polarimetric observations of selected field stars in the field of view of \gro.}
\label{polfs}
\begin{center}
\begin{tabular}{lccccc}
\noalign{\smallskip}    \hline\noalign{\smallskip}
ID      &JD             &PD          &EPVA              &$q$                    &$u$         \\
        &(2,400,000+)   &(\%)        &(deg.)            &                       &              \\
\hline\noalign{\smallskip}      
fs1     &56883.512      &$3.98\pm0.15$  &$68.3\pm1.1$   &$-0.0289\pm0.0018$     &$0.0273\pm0.0018$   \\
fs2     &57665.300      &$0.87\pm0.13$  &$49.6\pm4.4$   &$-0.0014\pm0.0014$     &$0.0085\pm0.0013$   \\
gfs2    &59496.319      &$5.38\pm0.45$  &$63.6\pm2.3$   &$-0.0326\pm0.0047$     &$0.0428\pm0.0048$    \\
\noalign{\smallskip}    \hline\noalign{\smallskip}
\end{tabular}
\end{center}
\end{table*}
%-----------------------------------

%-----------------------------------

\begin{longtable}{lcccccccc}
\caption{Results of the spectral analysis. Observations marked with a $\dag$
correspond to single-peak profiles.} \\
\label{specres} \\
\hline
\hline
Date            &JD             &\ew    &Err\_EW        &$\Delta V$     &Err\_$\Delta V$      &$\log(V/R)$    &Err\_$\log(V/R)$       &Telescope      \\
                &(2,400,000+)   &(\AA)  &(\AA)          &(km s$^{-1}$)  &(km s$^{-1}$)               &               &                       &       \\
\hline
\endfirsthead
\caption{continued}\\
\hline
Date            &JD             &\ew    &Err\_EW        &$\Delta V$     &Err\_$\Delta V$      &$\log(V/R)$    &Err\_$\log(V/R)$       &Telescope      \\
                &(2,400,000+)   &(\AA)  &(\AA)          &(km s$^{-1}$)  &(km s$^{-1}$)               &               &                       &       \\
\hline
\endhead
\hline
\endfoot
\hline
\endlastfoot
\hline
2004-06-25      &53182   &3.86   &0.30    &417   &15      &-0.074  &0.057  &SKO     \\
2004-07-06      &53193   &4.85   &0.34    &425   &19      &-0.036  &0.046  &SKO     \\
2004-08-25      &53243   &4.16   &0.23    &398   &13      &-0.067  &0.036  &SKO     \\
2004-09-13      &53262   &4.91   &0.19    &392   &9       &-0.019  &0.030  &SKO     \\
2004-10-03      &53282   &4.49   &0.27    &378   &15      &-0.046  &0.044  &SKO     \\
2004-10-24      &53303   &2.93   &0.25    &431   &20      &-0.034  &0.056  &SKO     \\
%2005-05-24     &53515  & 1.04   &0.19    &535   &40      &-0.350  &0.174  &SKO     \\
2005-06-22      &53544   &0.27   &0.16    &627   &32      &-0.099  &0.153  &SKO     \\
2005-07-12      &53564   &0.48   &0.10    &557   &18      &-0.036  &0.126  &SKO     \\
2005-07-13      &53565   &0.31   &0.11    &609   &20      &0.178   &0.128  &SKO     \\
2005-07-29      &53581   &1.19   &0.13    &572   &14      &0.042   &0.076  &SKO     \\
2005-08-16      &53599   &0.87   &0.09    &592   &20      &-0.207  &0.107  &SKO     \\
2005-09-20      &53634   &1.36   &0.14    &529   &20      &0.029   &0.087  &SKO     \\
2005-10-26      &53670   &1.77   &0.33    &527   &15      &-0.075  &0.079  &SKO     \\
2006-06-20      &53907   &2.48   &0.17    &458   &20      &-0.038  &0.110  &SKO     \\
2006-10-02      &54011   &3.21   &0.41    &499   &18      &0.002   &0.070  &SKO     \\
2006-10-24      &54033   &2.66   &0.15    &506   &14      &-0.113  &0.055  &SKO     \\
2007-05-14      &54235   &3.57   &0.24    &374   &70      &-0.076  &0.106  &SKO     \\
2007-05-29      &54250   &4.31   &0.21    &393   &15      &-0.069  &0.056  &SKO     \\
2007-09-05      &54349   &5.70   &0.19    &352   &10      &0.030   &0.026  &SKO     \\
%2007-09-09     &54353  & 7.60   &1.22    &420   &26      &-0.287  &0.149  &SKO     \\
2007-09-11      &54355   &6.17   &0.15    &355   &10      &-0.003  &0.026  &SKO     \\
2007-10-02      &54376   &5.03   &0.19    &370   &16      &-0.072  &0.040  &SKO     \\
2007-10-03      &54377   &5.07   &0.17    &365   &11      &-0.111  &0.035  &SKO     \\
2007-10-04      &54378   &5.13   &0.11    &386   &10      &-0.097  &0.029  &SKO     \\
%2008-05-12     &54599  & 6.55   &0.22    &318   &72      &-0.371  &0.202  &SKO     \\
2008-06-24      &54642   &7.35   &0.21    &345   &10      &-0.095  &0.022  &SKO     \\
2008-06-26      &54644   &7.28   &0.16    &326   &12      &-0.138  &0.028  &SKO     \\
%2008-07-14     &54662  & 7.64   &0.20    &271   &20      &0.672   &0.091  &SKO     \\
2008-07-15      &54663   &7.14   &0.20    &374   &27      &-0.139  &0.052  &SKO     \\
2008-08-12      &54691   &7.00   &0.18    &325   &15      &-0.140  &0.034  &SKO     \\
2008-09-02      &54712   &7.46   &0.18    &323   &16      &-0.041  &0.037  &SKO     \\
2009-07-29      &55042   &9.37   &0.16    &317   &8       &-0.015  &0.019  &SKO     \\
2009-08-11      &55055   &8.61   &1.00    &333   &19      &0.033   &0.051  &SKO     \\
2009-09-28      &55103   &10.02  &0.25    &322   &14      &-0.043  &0.030  &SKO     \\
2010-08-02      &55411   &7.67   &0.15    &351   &7       &0.047   &0.021  &SKO     \\
2010-08-27      &55436   &7.51   &0.17    &333   &8       &-0.008  &0.020  &SKO     \\
2010-09-14      &55454   &8.43   &0.16    &359   &8       &0.031   &0.019  &SKO     \\
2010-09-30      &55470   &8.47   &0.16    &316   &10      &-0.021  &0.025  &SKO     \\
2010-11-03      &55504   &7.24   &0.30    &360   &24      &-0.181  &0.064  &SKO     \\
2011-04-09      &55661   &7.46   &0.64    &302   &17      &0.051   &0.058  &LT      \\
2011-04-23      &55675   &7.84   &0.67    &332   &19      &0.031   &0.049  &LT      \\
2011-05-15      &55697   &6.22   &0.31    &293   &18      &-0.029  &0.127  &LT      \\
2011-05-20      &55702   &6.89   &0.51    &327   &20      &0.073   &0.054  &LT      \\
2011-06-12      &55725   &6.47   &0.37    &308   &51      &-0.074  &0.118  &LT      \\
2011-06-17      &55730   &7.61   &0.38    &356   &14      &-0.029  &0.041  &LT      \\
2011-06-25      &55738   &7.21   &0.22    &339   &10      &0.018   &0.040  &LT      \\
2011-07-02      &55745   &7.97   &0.34    &336   &18      &0.016   &0.056  &LT      \\
2011-07-08      &55751   &7.25   &0.43    &356   &14      &-0.119  &0.038  &LT      \\
2011-07-15      &55758   &6.62   &0.38    &388   &19      &0.087   &0.056  &LT      \\
2011-08-03      &55777   &7.50   &0.20    &368   &7       &0.012   &0.021  &SKO     \\
%2011-09-06     &55811  & 8.62   &0.96    &415   &28      &0.358   &0.128  &SKO     \\
2011-10-07      &55842   &6.42   &0.48    &353   &13      &-0.028  &0.054  &LT      \\
2011-10-14$\dag$&55849 &  4.29   &0.37    &--    &--      &--      &--     &LT       \\
%2011-10-29     &55864  & 14.55  &5.64    &0     &0       &9.999   &0.000  &LT      \\
2011-12-02      &55898   &5.20   &0.32    &362   &16      &0.015   &0.066  &LT      \\
2012-03-11      &55998   &3.44   &0.68    &454   &33      &0.058   &0.063  &LT      \\
2012-06-06      &56085   &3.85   &0.18    &420   &8       &0.014   &0.030  &SKO     \\
%2012-06-30     &56109  &5.30   &0.51    &0      &0       &9.999   &0.000  &LT      \\
%2012-08-03     &56143  &2.82   &0.35    &451   &46       &0.061   &0.140  &LT      \\
2012-08-24      &56164   &1.71   &0.07    &493   &14      &-0.100  &0.073  &SKO     \\
2012-08-25      &56165   &2.26   &0.23    &503   &16      &-0.188  &0.098  &SKO     \\
2012-09-02      &56173   &1.16   &0.25    &485   &18      &-0.111  &0.109  &LT      \\
2012-09-06      &56177   &2.21   &0.26    &485   &13      &-0.015  &0.070  &SKO     \\
2012-09-29      &56200   &1.26   &0.26    &578   &31      &-0.138  &0.125  &LT      \\
2012-10-19      &56220   &1.67   &0.15    &548   &14      &0.011   &0.068  &SKO     \\
2012-10-26      &56227   &1.68   &0.38    &570   &43      &-0.109  &0.145  &LT      \\
2013-05-17      &56430   &2.74   &0.53    &427   &34      &0.092   &0.148  &LT      \\
2013-06-21      &56465   &2.98   &0.49    &451   &37      &0.101   &0.118  &LT      \\
2013-06-28      &56472   &1.99   &0.23    &464   &25      &0.190   &0.102  &LT      \\
2013-07-19      &56493   &2.38   &0.23    &506   &18      &-0.013  &0.090  &LT      \\
2013-07-26      &56500   &3.53   &0.27    &459   &16      &0.004   &0.063  &LT      \\
2013-07-30      &56504   &3.18   &0.17    &452   &9       &-0.025  &0.042  &SKO     \\
2013-08-30      &56535   &3.35   &0.20    &475   &10      &0.014   &0.035  &SKO     \\
2013-10-18      &56584   &2.65   &0.11    &448   &8       &0.023   &0.042  &SKO     \\
2014-06-06      &56815   &4.56   &0.15    &395   &16      &0.073   &0.050  &SKO     \\
2014-08-05      &56875   &2.34   &0.14    &472   &11      &0.020   &0.048  &SKO     \\
2014-08-19      &56889   &2.28   &0.15    &500   &11      &-0.061  &0.052  &SKO     \\
2014-10-12      &56943   &2.33   &0.14    &496   &10      &-0.020  &0.052  &SKO     \\
2015-06-23      &57197   &4.29   &0.21    &374   &10      &0.058   &0.031  &SKO     \\
2015-07-08      &57212   &4.54   &0.24    &347   &9       &0.038   &0.027  &SKO     \\
2015-07-21      &57225   &5.01   &0.23    &355   &22      &0.002   &0.056  &SKO     \\
2015-10-05      &57301   &5.22   &0.22    &346   &12      &0.050   &0.040  &SKO     \\
2015-11-27      &57354   &5.52   &0.21    &353   &18      &0.050   &0.053  &LT      \\
2016-04-23      &57502   &7.14   &0.21    &310   &17      &0.042   &0.042  &LT      \\
2016-04-28      &57507   &7.73   &0.34    &307   &11      &-0.042  &0.029  &LT      \\
2016-05-15      &57524   &7.84   &0.37    &240   &29      &0.086   &0.085  &LT      \\
2016-05-21      &57530   &7.48   &0.25    &287   &28      &9.999   &0.000  &LT      \\
2016-06-08      &57548   &6.89   &0.10    &273   &78      &-0.056  &0.207  &SKO     \\
2016-07-12      &57582   &7.66   &0.26    &307   &15      &-0.047  &0.037  &LT      \\
2016-07-18      &57588   &7.52   &0.29    &316   &19      &0.005   &0.043  &LT      \\
2016-09-06      &57638   &6.79   &0.15    &405   &38      &-0.080  &0.066  &SKO     \\
2016-09-14      &57646   &7.97   &0.30    &272   &11      &0.001   &0.000  &LT      \\
2016-10-04      &57666   &6.52   &0.13    &333   &11      &0.129   &0.026  &SKO     \\
2016-10-14      &57676   &7.49   &0.20    &306   &13      &-0.045  &0.033  &LT      \\
2017-05-16      &57890   &8.10   &0.90    &398   &54      &-0.077  &0.088  &LT      \\
2017-05-28      &57902   &8.04   &0.83    &303   &15      &0.040   &0.038  &LT      \\
2017-06-03      &57908   &7.54   &0.84    &269   &45      &-0.010  &0.115  &LT      \\
2017-06-08      &57913   &9.45   &0.58    &262   &46      &0.130   &0.126  &LT      \\
2017-06-13      &57918   &8.24   &0.76    &248   &27      &0.048   &0.074  &LT      \\
2017-06-21      &57926   &10.24  &0.92    &264   &17      &-0.228  &0.044  &LT      \\
2017-06-26      &57931   &8.17   &0.84    &242   &20      &0.077   &0.060  &SKO     \\
2017-06-29      &57934   &9.09   &0.38    &268   &23      &-0.177  &0.062  &LT      \\
2017-07-13      &57948   &8.23   &0.20    &324   &8       &-0.040  &0.019  &SKO     \\
2017-08-13      &57979   &8.33   &0.32    &295   &27      &-0.060  &0.063  &LT      \\
2017-09-02      &57999   &8.10   &0.17    &270   &26      &-0.127  &0.066  &LT      \\
2017-09-17      &58014   &8.20   &0.32    &286   &17      &-0.034  &0.043  &LT      \\
2017-09-29      &58026   &7.72   &0.24    &339   &15      &0.047   &0.041  &LT      \\
%2017-10-11     &58038  &9.50   &0.41    &134   &43       &-0.266  &0.180  &LT      \\
2017-10-13      &58040   &8.77   &0.18    &281   &12      &0.065   &0.028  &SKO     \\
2018-04-05      &58214   &9.03   &0.31    &296   &27      &0.054   &0.066  &LT      \\
2018-04-29$\dag$&58238   &7.94   &0.35    &--    &--      &--      &--     &LT      \\
%2018-05-15     &58254  &10.40  &0.17    &0      &0       &9.999   &0.000  &LT      \\
2018-05-27      &58266   &10.10  &0.31    &292   &24      &-0.176  &0.065  &LT      \\
2018-06-06      &58276   &10.36  &0.18    &268   &25      &0.073   &0.059  &LT      \\
2018-08-18$\dag$&58349   &9.74   &0.19    &--    &--      &--      &--     &LT      \\
2018-08-23      &58354   &9.05   &0.16    &256   &12      &-0.045  &0.033  &SKO     \\
2018-08-25      &58356   &8.30   &0.17    &282   &10      &0.048   &0.026  &SKO     \\
2018-08-29      &58360   &8.96   &0.20    &274   &27      &0.019   &0.064  &LT      \\
2018-09-09      &58371   &8.82   &0.23    &288   &21      &0.057   &0.050  &LT      \\
2018-09-18      &58380   &8.46   &0.14    &309   &9       &-0.063  &0.021  &SKO     \\
2018-09-21      &58383   &9.00   &0.25    &282   &31      &-0.255  &0.096  &LT      \\
2018-10-08      &58400   &8.78   &0.16    &276   &10      &-0.032  &0.025  &SKO     \\
2019-04-17      &58591   &11.20  &0.30    &329   &24      &0.022   &0.051  &          \\
2019-04-27      &58601   &11.35  &0.43    &296   &29      &0.066   &0.070  &LT      \\
2019-05-08$\dag$&58612   &12.40  &0.31    &--    &--      &--      &--     &LT      \\
2019-05-19      &58623   &11.82  &0.24    &278   &27      &-0.215  &0.077  &LT      \\
2019-05-29      &58633   &10.90  &0.24    &322   &15      &-0.159  &0.039  &LT      \\
2019-06-13      &58648   &12.97  &0.34    &330   &21      &-0.041  &0.044  &LT      \\
2019-06-23      &58658   &12.71  &0.37    &360   &14      &-0.003  &0.037  &LT      \\
2019-07-19      &58684   &13.02  &0.33    &294   &18      &0.015   &0.040  &LT      \\
2019-07-29      &58694   &13.58  &0.27    &293   &7       &0.004   &0.017  &SKO     \\
2019-08-10      &58706   &13.82  &0.47    &289   &24      &-0.107  &0.060  &LT      \\
2019-08-19      &58715   &13.01  &0.28    &286   &12      &-0.197  &0.032  &SKO     \\
2019-09-09      &58736   &12.23  &0.18    &331   &7       &-0.035  &0.016  &SKO     \\
2019-09-24      &58751   &11.99  &0.22    &327   &7       &-0.020  &0.017  &SKO     \\
2019-10-09      &58766   &13.11  &0.40    &326   &8       &-0.001  &0.021  &LT      \\
%2019-12-22     &58840  &13.57  &0.89    &382   &101      &0.047   &0.188  &LT      \\
2020-06-22      &59023   &9.12   &0.26    &350   &6       &0.029   &0.015  &SKO     \\
%2020-07-02     &59033  &7.67   &0.58    &290   &77       &0.033   &0.141  &LT      \\
2020-07-20      &59051   &9.66   &0.21    &333   &5       &-0.049  &0.014  &SKO     \\
2020-08-05      &59067   &7.62   &0.13    &326   &70      &0.045   &0.126  &LT      \\
2020-08-25      &59087   &8.54   &0.21    &340   &8       &-0.050  &0.022  &SKO     \\
2020-09-14      &59107   &8.63   &0.23    &330   &7       &-0.037  &0.018  &SKO     \\
2020-09-16      &59109   &8.72   &0.29    &371   &30      &-0.094  &0.066  &SKO     \\
2020-09-29      &59122   &8.65   &0.23    &322   &10      &0.006   &0.023  &SKO     \\
2020-09-30      &59123   &8.27   &0.49    &336   &14      &-0.077  &0.057  &LT      \\
2020-10-30      &59153   &8.74   &0.35    &346   &14      &-0.027  &0.050  &LT      \\
2021-07-02      &59398   &6.96   &0.20    &388   &6       &-0.052  &0.017  &SKO     \\
2021-08-11      &59438   &6.43   &0.28    &348   &7       &-0.130  &0.022  &SKO     \\
2021-09-01      &59459   &6.35   &0.19    &370   &4       &-0.005  &0.017  &SKO     \\
2021-09-05      &59463   &6.65   &0.25    &377   &5       &-0.008  &0.019  &SKO     \\
2021-09-29      &59487   &6.59   &0.20    &377   &6       &-0.056  &0.020  &SKO     \\
2022-05-24      &59724   &3.26   &0.29    &420   &9       &0.065   &0.035  &SKO     \\
2022-06-15      &59746   &2.72   &0.20    &442   &8       &-0.021  &0.039  &SKO     \\
2022-07-04      &59765   &3.26   &0.10    &442   &8       &-0.144  &0.037  &SKO     \\
2022-07-06      &59767   &2.82   &0.18    &415   &11      &0.005   &0.045  &SKO     \\
2022-07-30      &59791   &1.35   &0.17    &502   &12      &0.032   &0.061  &SKO     \\
2022-09-08      &59831   &1.15   &0.12    &532   &11      &0.158   &0.057  &SKO     \\
2022-09-09      &59832   &1.75   &0.21    &530   &11      &0.058   &0.053  &SKO     \\
2022-09-19      &59842   &1.52   &0.12    &505   &9       &-0.004  &0.047  &SKO     \\

\hline
\end{longtable}

%-----------------------------------------------------------------------------------------

\end{appendix}

\end{document}